\documentclass[structabstract]{aa}  
\usepackage{graphicx}
\usepackage{rotating}
\usepackage{pdflscape}
\usepackage{subfigure}
\usepackage{longtable}
\usepackage{txfonts}

\usepackage{natbib}
\usepackage{enumerate}

\bibpunct{(}{)}{;}{a}{}{,} 

\begin{document}

   \title{The Gaia-ESO Survey: Abundance ratios in the inner-disk open clusters Trumpler~20, NGC~4815, NGC~6705\thanks{Based on observations collected with the FLAMES  spectrograph at VLT/UT2 telescope (Paranal Observatory, ESO, Chile), for the Gaia-ESO Large Public Survey (188.B-3002).}}


\author{L. Magrini\inst{1}, 
S. Randich\inst{1}, 
D. Romano\inst{2}, 
E. Friel\inst{3},  
A. Bragaglia\inst{2}, 
R. Smiljanic\inst{4,5}, 
H. Jacobson\inst{6}, 
A. Vallenari\inst{7},
M. Tosi\inst{2},  
L. Spina\inst{1,8},  
P. Donati\inst{2,9}, 
E. Maiorca\inst{1}, 
T. Cantat-Gaudin\inst{7,10}, 
R. Sordo\inst{7}, 
M. Bergemann\inst{11}, 
F. Damiani\inst{12},
G.  Tautvai\v{s}ien\.{e}\inst{13},
S. Blanco-Cuaresma\inst{14, 15},
F. Jim{\'e}nez-Esteban\inst{16},
D. Geisler\inst{17},
N. Mowlavi\inst{18},
C. Munoz\inst{17},
I. San Roman\inst{17},
C. Soubiran\inst{15},
S. Villanova\inst{17},
S. Zaggia\inst{7},
G. Gilmore\inst{11},
M. Asplund\inst{19},
S. Feltzing\inst{20}, 
R. Jeffries\inst{22}, 
T. Bensby\inst{20}, 
P. Francois\inst{21}, 
S. Koposov\inst{11,23}, 
A. J. Korn\inst{20}, 
E. Flaccomio\inst{12}, 
E. Pancino\inst{2,24}, 
A. Recio-Blanco\inst{25},  
G. Sacco\inst{1}, 
M. T.  Costado\inst{26},
E. Franciosini\inst{1}, 
P. Jofre\inst{11}, 
P. de Laverny\inst{25}, 
V. Hill\inst{25}, 
U. Heiter\inst{27}, 
A. Hourihane\inst{11}, 
R. Jackson\inst{22}, 
C. Lardo\inst{2}, 
L. Morbidelli\inst{1}, 
J. Lewis\inst{11}, 
K. Lind\inst{11}, 
T. Masseron\inst{11},
L. Prisinzano\inst{12},
C. Worley\inst{11}
}
\offprints{L. Magrini}

\institute{
INAF--Osservatorio Astrofisico di Arcetri, Largo E. Fermi, 5, I-50125 Firenze, Italy
\email{laura@arcetri.astro.it}
\and
INAF-Osservatorio Astronomico di Bologna, Via Ranzani, 1, 40127, Bologna, Italy
\and
Department of Astronomy, Indiana University, Bloomington,  USA
\and
Department for Astrophysics, Nicolaus Copernicus Astronomical Center, ul. Rabia\'nska 8, 87-100 Toru\'n, Poland
\and
European Southern Observatory, Karl-Schwarzschild-Str. 2, 85748 Garching bei M\"unchen, Germany
\and
MIT Kavli Institute, Boston, USA
\and
Osservatorio Astronomico di Padova, Vicolo dell'Osservatorio, 5, 35122, Padova, Italy
\and
Dipartimento di Fisica, sezione di Astronomia, Largo E. Fermi, 2, I-50125 Firenze, Italy
\and
Dipartimento di Fisica e Astronomia,  Via Ranzani, 1, 40127, Bologna, Italy
\and
Dipartimento di Fisica e Astronomia, Vicolo dell'Osservatorio, 3, 35122, Padova, Italy
\and
Institute of Astronomy, University of Cambridge, Madingley Road, Cambridge, CB3 0HA, UK 
\and
Osservatorio Astronomico di Palermo, 
P.zza del Parlamento 1, 90134, Palermo, Italy
\and
Institute of Theoretical Physics and Astronomy, Vilnius University, A. Gostauto 12, 01108 Vilnius, Lithuania 
\and
Univ. Bordeaux, LAB, UMR 5804, F-33270, Floirac, France.
\and
CNRS, LAB, UMR 5804, F-33270, Floirac, France
\and 
Centro de Astrobiolog{\'i}a (INTA-CSIC)
P.O. Box 78, 28691 Villanueva de la Canada,  Madrid, Spain
\and
Departamento de Astronom'a, Universidad de Concepci{\'o}n, Casilla 160-C, Concepci{\'o}n, Chile
\and
Astronomy Department, University of Geneva, Ch. des Maillettes 51, 1290 Versoix, Switzerland
\and
Research School of Astronomy and Astrophysics, Australian National University, Canberra, ACT 2611, Australia
\and 
Lund Observatory, Department of Astronomy and Theoretical Physics, Box 43, SE-221\,00 Lund, Sweden
\and
Observatoire de Paris, CNRS, Univ. Paris Diderot, place Jules Janssen, 92190, Meudon, France
\and
Astrophysics Group, Keele University, Keele, Staffordshire, ST5 5BG, UK
\and
Moscow M.V. Lomonosov State University, Sternberg Astronomical Institute, Universitetskij pr., 13, 119992 Moscow, Russia
\and 
ASI Science Data Center, I-00044 Frascati, Italy 
\and
Laboratoire Lagrange (UMR7293), UniversitŽ de Nice Sophia Antipolis, CNRS,
Observatoire de la C™te dÕAzur, BP 4229,F-06304 Nice cedex 4, France
\and
Instituto de Astrof{\'i}sica de Andaluc{\'i}a (IAA-CSIC), Glorieta de la Astronom{\'i}a, E-18008-Granada, Spain
\and
Department of Physics and Astronomy, Uppsala University, Box 516, 75120 Uppsala, Sweden }
   \date{Received ; accepted }

 
  \abstract
{Open clusters are key tools to study the spatial distribution of abundances in the disk and their evolution 
with time.  }
 {Using the first release of stellar parameters and abundances of the Gaia-ESO Survey, we analyse the chemical properties of stars in three old/intermediate-age open clusters, namely NGC~6705, NGC~4815, and Trumpler~20, all located in 
the inner part of the Galactic disk at Galactocentric radius R$_{GC}\sim$7~kpc, 
aiming at proving their homogeneity and at comparing them with the field population.    }
{We study the abundance ratios of elements belonging to two different nucleosynthetic channels:  $\alpha$-elements and  iron-peak elements. 
For each element we analyse the internal chemical homogeneity of cluster members and we compare the cumulative distributions of cluster abundance ratios with those of solar neighbourhood turn-off stars and of inner-disk/bulge giants.   We compare the  abundance ratios of field and cluster stars with two chemical evolution models  that predict 
different  $\alpha$-enhancement dependences on the Galactocentric 
distance due to different assumptions on the infall and star formation rates.  }
 {The main results can be summarised as follows: 
i) cluster members are chemically homogeneous within 3-$\sigma$ in all analysed elements; ii)   the three clusters have comparable [El/Fe] patters within $\sim$1-$\sigma$, but they differ in their global metal content [El/H], with NGC~4815 having the lowest metallicity. Their 
[El/Fe] ratios show differences and analogies with those of the field population, both in 
the solar neighbourhood and in the bulge/inner disk; 
iii)  comparing the abundance ratios with  the results of two
chemical evolution models and with field star abundance distributions, we find that the 
abundance ratios of Mg, Ni, Ca  in NGC~6705 might require an inner birthplace, implying a subsequent variation of its R$_{GC}$ during its lifetime, 
consistent with previous orbit determination.   
}
 {Using the results of the first internal data release, we show the potential of  the Gaia-ESO Survey, 
through  a homogeneous and detailed analysis of the cluster versus  field populations, to reveal 
the chemical structure of our Galaxy using a completely uniform analysis of different populations. 
We verify  that the Gaia-ESO Survey data  are able to identify the unique chemical properties of each cluster, pinpointing the composition of the interstellar medium at the epoch and place of formation. 
The full dataset of the Gaia-ESO Survey  will be a superlative tool to constrain the chemical evolution of our Galaxy by disentangling  different formation and evolution scenarios. }
\authorrunning{Magrini, L. et al.}
\titlerunning{\sc Abundance ratios in the Gaia-ESO Survey old clusters}
\maketitle

\section{Introduction}

Open clusters are very useful tracers of  the processes of formation and evolution of our Galaxy. 
They are a  disc population located from the inner parts of the disk to 
its outskirts, and with ages spanning from few Myr for recently formed clusters  to several Gyr for old clusters (see, e.g., Dias et al., 2002; Dias et al., 2012).   
The population of young open clusters has a small scale height above the Galactic plane ($\sim$60~pc), while old open clusters 
reach higher altitudes $\sim$350~pc \citep{chen03}, thus being all presumably part of the thin component of the disk. 
Kinematics of open clusters, both in terms of rotation and velocity dispersions, are also consistent with association with the thin disk of the Galaxy 
\citep{scott95, wu09}. In addition, also form a chemical point of view, the roughly solar abundance ratios of open clusters further supports to an association with the thin disk. 

Their ages and distances  can be derived from colour-magnitude diagrams, obtained via photometric studies, making them 
a perfect instrument to investigate the temporal changes in the spatial distribution of abundances  (e.g., Magrini et al. 2009, Yong et al., 2012). 

In the framework of the study of  cluster population, the Gaia-ESO Survey (Gilmore et al. 2012, Randich \& Gilmore 2012), together with other spectroscopic surveys observing open clusters, such as the Apache Point Observatory Galactic Evolution Experiment (APOGEE, Allende-Prieto et al. 2008),  is allowing us to study a large number of young, intermediate-age, and old open clusters.  Within the APOGEE project, the Open Cluster Chemical Analysis and Mapping (OCCAM) survey aims to produce a comprehensive, uniform, infrared-based data set for hundreds of open clusters, and constrain key Galactic dynamical and chemical parameters from this sample. A first contribution to the OCCAM project has been recently published by \citet{frinchaboy13} that presented the 
analysis of 141 members stars in 28 open clusters. This new dataset allowed them to revise the Galactic metallicity gradient.

In particular, old and intermediate-age clusters are
rare fossils of the past star formation history of the Galactic disk. 
Apparently, the old open clusters observed at present time with ages over $\sim$1 Gyr have survived 
because of their peculiar initial characteristics, such as their larger than average mass, 
higher central concentration, and orbits that allow them to avoid the disruptive influence of the giant 
molecular clouds \citep{friel95, janes94, bonatto06}.
This might introduce possible differences in chemical composition between field and cluster stars presently 
observed at the same Galactocentric radius, since both might have moved from their place of birth, 
but in different ways.

The majority of stars born in open clusters were indeed dispersed into the Galaxy field in a relatively short time, i.e., within the first Gyr from their formation \citep[see, e.g.,][]{janes94, lada03, gieles06}.  This phenomenon is  more rapid in the inner disk where the density of stars is higher \citep{freeman70, vanderkruit02}.
Thus the existence of several old and intermediate-age open clusters within the Solar circle (R$_{\rm GC}<$8~kpc), such as those observed during the first 
periods of observation of the Gaia-ESO Survey,  offers a unique opportunity to study the evolution of the disk in a region so far little explored. 

Specifically, this study focuses on  detailed and homogeneously obtained  chemical abundance patterns of different populations, that offer 
valuable clues in the interpretation of the Galactic history. We refer to the paper of Jacobson et al. (in prep.) for a discussion of the spatial 
distribution of the global metallicity and its implications for the radial Galactic metallicity gradient. 

As a  first approximation, we expect  the abundances of open clusters to match those of the field stars at similar Galactocentric distances, but 
observations are revealing differences \citep[see, e.g.,][]{yong05, yong12,desilva07} and these differences contain important information about, e.g., the place 
where the open clusters were born, the homogeneity of the  disk at any R$_{\rm GC}$ at the epoch when the cluster formed, etc..

More in details, studies that compare field and cluster populations were tackled by several authors in the past.
For example, \citet{desilva07} analysed the chemical pattern of the inner old open cluster Collinder~261, comparing its abundance ratios  with 
those of Cepheid stars and of field stars. They found remarkable differences for some elements, such as Na, Mg, Si, and Ba. 
They claimed that the differences are a signature of the local inhomogeneities at the time and site of cluster  formation. 
Other examples can be found in the works of  \citet{yong05, yong12} and of \citet{sestito08}, who studied  open clusters located at R$_{\rm GC}>$13~kpc.  
They compared the chemical properties of several outer disk open clusters  with those of stellar clusters located in the disk within  R$_{\rm GC}<$13~kpc,
with other stellar tracers located in the outer disk, such as red giant stars, Cepheids, and the solar neighbourhood  stars. 
They found that the behaviour of $\alpha$-elements is not exactly the same in all their clusters, but that it is on average similar to that of solar neighbourhood stars. 
They concluded that the primary difference between solar neighbourhood and outer disk is that the chemical enrichment in the outer disk did not yet reach the metallicities of the solar neighbourhood, but the contribution of the two nucleosynthetic channels, SNII and SNIa, appear to have been similar.  
However, the conclusions of these works  are  usually based on heterogeneous samples including literature and authors' own results. 
Heterogeneous samples might mask genuine abundance differences and/or 
artificially induce (or amplify) abundance differences between otherwise
chemically similar populations. The different effects are driven by the size of the samples and by  the elements considered. 

In this framework the Gaia-ESO Survey data, with its uniform data-set and its homogeneous analysis, will allow, for the first time, a comparison of different populations on a 
footing not possible before.  

As an initial step in this direction, the first data release of the Gaia-ESO Survey, including the first six months of observations, allows us to analyse in  detail 
the chemical composition of three old and intermediate-age open clusters located in the very inner disk (NGC~6705, NGC~4815, and Trumpler~20), 
and to compare them with the field population in the solar neighbourhood  and with evolved stars located in the inner-disk/bulge.

The present  paper is structured as follows: 
In Sec.~2 we briefly  describe the Gaia-ESO Survey. In Sec.3 we present the properties and membership of the first three old/intermediate-age clusters observed by the Gaia-ESO Survey. In Sec.~4 
we summarise the target selection strategy for field stars. 
In Sec.~5  we check the quality of the analysis of cluster stars and  we analyse the abundance patterns of the three open cluster and compare them to the field population, whereas in Sec.~6 we speculate on the origin of their abundance ratios. 
In Sec.~7 we give our summary. 

\section{The Gaia-ESO Survey and its first data release} 
\label{sec_gaiaeso}
The Gaia-ESO Survey 
is a large, public spectroscopic survey started at the end of 2011 that is employing 
the VLT FLAMES \citep{pasquini02}  instrument to obtain high quality spectroscopy of $\sim$10$^5$ stars in our Galaxy. 
The observed stars belong to well defined samples  and are selected making use of several photometric databases such as   
the VISTA Hemisphere Survey (VHS) \citep{mcmahon12}, the Two Micron All Sky Survey (2MASS, Skrutskie et al. 2006) and a variety of photometric surveys of open clusters. 
The focus of the Gaia-ESO Survey  is to quantify the kinematical and chemical element abundance distributions 
in the different components of the Milky Way: bulge, thin and thick disks, halo, and about a  hundred  open clusters spanning a large range of 
ages, distances, and masses.  A general description  of the Survey can be found in Gilmore et al. (in prep.) and Randich et al. (in prep.).

In  the present work we discuss results of the analysis of UVES  \citep{dekker00} spectra of F-G-K stars. This analysis is described in details in Smiljanic et al. (in prep.).
We briefly review how recommended parameters are computed.  The recommended parameters are obtained combining the results of different nodes considering first 
the accuracy of  each node judged using a sample of calibration stars with well-known stellar parameters, called 
benchmark 
stars, as reference  (see \citet{jofre13a, jofre13b}).
Among the benchmark stars, there are stars having stellar
parameters comparable to that of the stars discussed in the present paper, as, e.g., $\xi$ Hya.

The different approaches of the nodes con be summarised as follows: {\em i)} nodes that employ the  equivalent 
width (EW) analysis, obtaining EWs form the observed spectra. The atmospheric parameter determination is based on the
excitation and ionisation balance of iron lines;
{\em ii)}  spectrum synthesis methods  that  derive the atmospheric parameters from a $\chi ^2$ fit to observed spectra. 
In some cases the computation of EWs from best-matching synthetic spectra is used to derive the individual element 
abundances; 
{\em iii)}  multi-linear regression methods that simultaneously
determine the stellar parameters  of an observed
spectrum by the projection of the spectrum onto
vector functions, constructed as an optimal
linear combination of the local synthetic spectra.
 
The nodes using the EW method are eight, while the remaining five nodes adopt different approaches to the spectral synthesis. Details on the individual node techniques are in Smiljanic et al. (in prep.).

Further consistency tests are conducted using the calibration
clusters and other calibration targets, such as, e.g., the globular clusters
NGC1851 and NGC2808, for which the nodes' and recommended parameters were compared 
 with PARSEC isochrones in the T$_{eff}$ vs. log g plane finding a good agreement, and  the CoRoT (COnvection ROtation and planetary Transits, Baglin et al. 2006) giant stars.
Then the parts of the parameter space where a given node under-performs are identified\footnote{Nodes that 
cannot analyse the benchmark stars and reproduce their atmospheric parameters
(T$_{eff}$ and log~g) within 150~K and 0.30~dex, respectively,  are disregarded. } 

Finally, the median value of the validated results is adopted as the
recommended value of that parameter. 
 For the cluster stars  discussed in the present paper, the recommended values are typically derived using the results of 8-10 nodes. For the solar neighbourhood stars and the inner-disk/bulge stars the parameters are, on average, determined using the results of nine nodes.

The uncertainties are taken to be the method-to-method dispersions.
Once the recommended values of the atmospheric parameters
of all stars are defined, the spectroscopic analysis proceeds
to its second step, the determination of elemental abundances.
The nodes use the recommended parameters to re-compute elemental abundances, and 
the median values of the resulting abundances are the final recommended best values.

{\em We emphasise  here that cluster and field stars are analysed in a completely homogeneous way}. Parameters 
and abundances for stars observed during the first six months have been delivered within the Gaia-ESO Survey consortium in an internal Data Release (GESviDR1Final), 
which includes abundances   of FeI, FeII, NiI, CrI, TiI, TiII, SiI, CaI, MgI, NaI, AlI, ZnI, YII, ZrII,  CeII. Unfortunately, the abundances of neutron capture elements 
are available only for a small sub-sample of stars,  and thus they are not discussed in the present paper.
The recommended values 
are used in 
the rest of the paper. In particular, we use the abundances derived with the recommended parameters, including those of iron
 (computed considering only Fe~I lines). The Gaia-ESO survey abundances are scaled to the solar abundances
of \citet{grevesse07}. 
The errors on [Fe/H] are comparable among the different samples of solar-neighbourhood stars, inner disk stars, and intermediate-age cluster stars discussed in the present paper, with cool stars having in general higher errors  both in cluster and Milky Way stars. 
Typical errors are of the order of 0.1~dex, and they become $\sim$0.2 dex for several stars cooler than $\sim$4500 K.

\section{The old and intermediate-age clusters in DR1}
Three old and intermediate-age open clusters were observed and analysed in DR~1: 
NGC~6705, NGC4815, and Trumpler~20. 
They are all located within the Solar circle  in a region still poorly investigated but 
of great importance for our understanding of the mechanisms of disk/bulge formation.  
Being close to the Galactic centre (see Table~\ref{tab_par}), the three clusters might suffer from strong tidal effects, 
as well as frequent interactions with molecular clouds, and thus they can provide important constraints to the cluster survival in a 'hostile' environment
\citep[see, e.g.,][]{janes94, lada03, gieles06}.
They might also probe a key issue to understand the mechanism of Galaxy formation and evolution, i.e., the radial metallicity gradient, 
which was the subject of a number of studies in past decades using open clusters as tracers \citep[see, e.g.,][]{janes79, panagia81, friel95, twarog97, carraro98, bragaglia06, sestito06, sestito07, 
sestito08, magrini09, pancino10, andreuzzi11, jacobson11}.   
The first results from the Gaia-ESO Survey are helping to shed light on the radial metallicity gradient in the very inner part of the disk, a region 
still poorly investigated (Jacobson et al. in prep.). 
Stars observed in the three clusters were selected on the basis of their colour-magnitude diagrams: targets for GIRAFFE were mainly main-sequence stars, 
while targets for UVES were evolved stars belonging to the red clump. More details on the target selection can be found in Bragaglia et al. (in prep.) and the papers 
on individual clusters, as, e.g., Donati et al. (in prep.) for Trumpler 20 and Friel et al. (in prep.) for NGC~4815. 

In the present paper, we show the results of [Fe/H] and abundance ratios of  member stars of the three  clusters presented above  and observed with UVES. 
The membership has been derived using the information on the radial velocities measured in  the Gaia-ESO Survey spectra  in the paper of 
Donati et al. (in prep) for Trumpler~20, in Friel et al. (in prep.) for NGC~4815, and in Cantat-Gaudin et al. (in prep.) for NGC~6705. 
Briefly, the radial velocities  are firstly used to identify the systemic velocity of the cluster, and then to remove stars beyond a certain $\sigma$ level from the median velocity
to separate cluster member stars from the field. 

The cluster parameters are summarised in Table~\ref{tab_par}, where we show the names of the clusters, their coordinates (equatorial and Galactic), the reddening E(B-V), the age in Gyr, the turn-off (TO) mass in solar masses, the distance from the Sun, the Galactocentric distance, and the height from the Galactic plane 
in kpc,  the metallicity, [Fe/H], and the number of members analysed. 
For a discussion and comparison with literature values of the three clusters we refer to the papers of Donati et al. (2013), Friel et al. (in prep.), and Cantat-Gaudin et al. (in prep.).

\begin{table*}
\begin{center}
\caption{Clusters' parameters.  }
\tiny
\begin{tabular}{lllllllllllc}
\hline\hline
Name & $\alpha$   &    $\delta$ &  l   		& b 		& E(B-V) & Age    	 & TO mass      & D$_{\odot}$  & R$_{GC}$$^d$  & [Fe/H] & N.of members\\
	  & \multicolumn{2}{c}{J2000.0}	& (deg)		& (deg)		&		&(Gyr)   & (M$_{\odot}$) & (kpc)       & (kpc) & &\\
\hline\hline
NGC6705$^a$  	& 18:51:05 &	-06:16:12  & 27.31    & -2.78  & 0.43  & 0.30$\pm$0.05		      	             & $\sim$3       &  1.9             & 6.3               & +0.14$\pm$0.06  & 21\\
Trumpler20$^b$	&12:39:32  &     -60:37:36  & 301.48 & 2.22   & 0.33  & 1.50$\pm$0.15  	    &$\sim$1.8     &   2.4             & 6.88     & +0.17$\pm$0.05& 13\\
NGC4815$^c$  	&12:57:59  &	-64:57:36   & 303.63 &-2.10  & 0.72  & 0.57$\pm$0.07 			       &$\sim$2.5     &   2.5             & 6.9               &  +0.03$\pm$0.05& 5\\
\hline \hline
\end{tabular}
\label{tab_par}\\
\end{center}
{\it a}-- NGC~6705 parameters from Cantat-Gaudin et al. (in prep.). 
{\it b}-- Trumpler~20 parameters from Donati et al. (in prep.). 
{\it c}-- NGC~4815 parameters from Friel et al. (in prep.). 
{\it d}-- computed with R$_{\odot}$=8~kpc. 
\end{table*}

\section{UVES observations of field stars in DR1}

To compare the abundance ratios of the different populations of the Milky Way and of open clusters  in a fully homogeneous way, we have  considered the results in GESviDR1Final for the Milky Way field stars.
These stars belong to two samples: the solar neighbourhood sample and the bulge/inner disk sample. 
For a complete description about how these samples are defined we refer to Gilmore et al. (in prep.). 
\paragraph{The solar neighbourhood sample} 
  The  observation criteria for the UVES stars aim at including the three major solar neighbourhood
  population groups (halo, thick disk, old thin disk). 
 They are designed to obtain an unbiased sample of $\sim$5000 G-stars within  2~kpc from the Sun. 
 The purpose of this sample is to quantify in detail the local elemental abundance distribution functions. 
 The sample used for the present work includes 390 stars, and corresponds 
to the Milky Way turn-off (TO) stars with recommended parameters in  GESviDR1Final.
   \paragraph{The bulge and inner disk}
The UVES stars observed together with the GIRAFFE stars dedicated to the study of 
Galactic bulge are expected to be evolved stars belonging to both bulge and inner disk populations. 
The prime targets of the GIRAFFE observations are K giants, including
the red clump  stars, with  typical magnitude $I=$15. 
The brighter K giant stars in the same fields are the targets of UVES, and they allow sampling
bulge and inner Galaxy populations. 
During the first six months of the  Gaia-ESO Survey,  32 stars of the bulge/inner-disk were observed and are included in the 
present discussion.  

\section{The abundance patterns of open clusters} 

The first step of our analysis is thus to check if the abundance ratios in each clusters are independent of the stellar parameters. 
To derive the abundance ratios we use the iron abundance computed by the nodes with the recommended stellar parameters. 
In Table~\ref{tab_dr1} we present the ids, the radial velocities, the recommended values of stellar parameters (T$_{eff}$,  log~g    and $\xi$) with their errors and the elemental abundance 
with their error (expressed in the logarithmic form A(El)=12+$\log$(El/H)) 
for  members in  open clusters.  As anticipated in Sec.~\ref{sec_gaiaeso}, the uncertainties on the stellar parameters (T$_{eff}$,  log~g, $\xi$ and [Fe/H]) in DR~1 are the method-to-method dispersions. 
For Fe~I two errors are given: the method-to-method dispersion   ($\sigma$ (FeI), in column~6) computed in the calculation of the recommended atmospheric parameters, 
and the node-to-node dispersion in A(FeI) of abundances re-computed with the recommended parameters (the error on [Fe/H] presented in column~5).

\begin{landscape}
\begin{table}
\scriptsize
\caption{Stellar parameters and abundance for member stars. }
\begin{tabular}{llllllllllllll}
\hline \hline
  Id           &  T$_{eff}$    & log~g          &  $\xi$         & A(FeI)          &  $\sigma$ (FeI)  &A(FeII)       &   A(MgI)         &  A(SiI)         &   A(CaI)          &  A(TiI)         &   A(CrI)         &   A(NiI)         \\
                &  K    &          &   km~s$^{-1}$          &           &    &     &            &         &            &          &           &          \\
\hline \hline																	        
  12391577-6034406   &  4849$\pm$41   &  2.86$\pm$0.13   &  1.27$\pm$0.11 &  7.50$\pm$0.03  &  0.06 &  7.46$\pm$0.05 &   7.67$\pm$0.09  &  7.54$\pm$0.05 &   6.38$\pm$0.05  &  4.91$\pm$0.02 &   5.66$\pm$0.07 &   6.18$\pm$0.03   \\
  12392585-6038279   &  5034$\pm$106  &  3.12$\pm$0.31   &  1.25$\pm$0.1  &  7.66$\pm$0.01  &  0.08 &  7.64$\pm$0.07 &   7.79$\pm$0.06  &  7.70$\pm$0.04 &   6.46$\pm$0.01  &  5.04$\pm$0.02 &   5.81$\pm$0.03 &   6.37$\pm$0.01  \\
  12392700-6036053  &  4800$\pm$77   &  2.80$\pm$0.24   &  1.29$\pm$0.1  &  7.60$\pm$0.03  &  0.07 &  7.57$\pm$0.04 &   7.72$\pm$0.02  &  7.63$\pm$0.06 &   6.38$\pm$0.04  &  4.91$\pm$0.03 &   5.67$\pm$0.02 &   6.34$\pm$0.04   \\
  12393132-6039422   &  4954$\pm$64   &  3.07$\pm$0.15   &  1.20$\pm$0.22 &  7.64$\pm$0.02  &  0.05 &  7.67$\pm$0.06 &   7.75$\pm$0.02  &  7.68$\pm$0.04 &   6.47$\pm$0.05  &  5.01$\pm$0.03 &   5.76$\pm$0.06 &   6.37$\pm$0.02  \\
  12393782-6039051   &  4909$\pm$129  &  2.80$\pm$0.21   &  1.30$\pm$0.16 &  7.61$\pm$0.04  &  0.12 &  7.62$\pm$0.06 &   7.75$\pm$0.06  &  7.66$\pm$0.04 &   6.42$\pm$0.04  &  4.97$\pm$0.01 &   5.71$\pm$0.02 &   6.32$\pm$0.02   \\
  12394419-6034412   &  4941$\pm$90   &  2.88$\pm$0.23   &  1.25$\pm$0.06 &  7.66$\pm$0.04  &  0.07 &  7.67$\pm$0.03 &   7.75$\pm$0.04  &  7.64$\pm$0.05 &   6.50$\pm$0.03  &  5.03$\pm$0.02 &   5.76$\pm$0.03 &   6.38$\pm$0.01  \\
  12394475-6038339   &  4850$\pm$112  &  2.75$\pm$0.22   &  1.38$\pm$0.08 &  7.58$\pm$0.02  &  0.05 &  7.62$\pm$0.06 &   7.75$\pm$0.04  &  7.66$\pm$0.02 &   6.39$\pm$0.02  &  4.93$\pm$0.04 &   5.71$\pm$0.04 &   6.32$\pm$0.02   \\
  12394596-6038389   &  4912$\pm$118  &  2.87$\pm$0.21   &  1.27$\pm$0.15 &  7.63$\pm$0.03  &  0.06 &  7.64$\pm$0.04 &   7.72$\pm$0.05  &  7.63$\pm$0.04 &   6.44$\pm$0.07  &  4.94$\pm$0.02 &   5.71$\pm$0.01 &   6.34$\pm$0.04  \\
  12394690-6033540   &  4968$\pm$77   &  3.03$\pm$0.1    &  1.14$\pm$0.14 &  7.70$\pm$0.04  &  0.11 &  7.69$\pm$0.04 &   7.79$\pm$0.04  &  7.70$\pm$0.05 &   6.47$\pm$0.04  &  5.01$\pm$0.01 &   5.76$\pm$0.03 &   6.42$\pm$0.02   \\
  12394742-6038411   &  4900$\pm$100  &  2.73$\pm$0.23   &  1.21$\pm$0.15 &  7.62$\pm$0.04  &  0.05 &  7.59$\pm$0.02 &   7.75$\pm$0.04  &  7.70$\pm$0.02 &   6.43$\pm$0.04  &  4.94$\pm$0.01 &   5.73$\pm$0.03 &   6.32$\pm$0.03  \\
  12395426-6038369   &  4925$\pm$100  &  2.98$\pm$0.13   &  1.36$\pm$0.04 &  7.63$\pm$0.02  &  0.07 &  7.66$\pm$0.08 &   7.74$\pm$0.05  &  7.67$\pm$0.04 &   6.42$\pm$0.03  &  5.01$\pm$0.02 &   5.72$\pm$0.06 &   6.37$\pm$0.02   \\
  12395975-6035072   &  4850$\pm$87   &  2.79$\pm$0.19   &  1.29$\pm$0.11 &  7.62$\pm$0.03  &  0.07 &  7.61$\pm$0.08 &   7.74$\pm$0.06  &  7.67$\pm$0.03 &   6.42$\pm$0.03  &  4.92$\pm$0.03 &   5.69$\pm$0.03 &   6.31$\pm$0.02  \\
  12400278-6041192   &  4932$\pm$67   &  2.98$\pm$0.11   &  1.37$\pm$0.05 &  7.58$\pm$0.02  &  0.10 &  7.61$\pm$0.07 &   7.71$\pm$0.07  &  7.66$\pm$0.03 &   6.42$\pm$0.03  &  4.96$\pm$0.03 &   5.71$\pm$0.07 &   6.32$\pm$0.04   \\
 \hline		   											        
  12572442-6455173   &  4198$\pm$88   &  1.56$\pm$0.37   &  1.37$\pm$0.05 &  7.48$\pm$0.03  &  0.07 &  7.53$\pm$0.03 &   7.85$\pm$0.04  &  7.6 $\pm$0.1  &   6.23$\pm$0.07  &  4.66$\pm$0.03 &   5.5 $\pm$0.01 &   6.2 $\pm$0.04  \\
  12574328-6457386   &  4895$\pm$40   &  2.40$\pm$0.13   &  1.73$\pm$0.14 &  7.45$\pm$0.02  &  0.08 &  7.46$\pm$0.05 &   7.68$\pm$0.1   &  7.47$\pm$0.05 &   6.29$\pm$0.05  &  4.75$\pm$0.03 &   5.54$\pm$0.03 &   6.13$\pm$0.04   \\
  12575511-6458483   &  4870$\pm$84   &  2.55$\pm$0.47   &  1.43$\pm$0.24 &  7.51$\pm$0.01  &  0.07 &  7.55$\pm$0.06 &   7.78$\pm$0.07  &  7.5 $\pm$0.01 &   6.32$\pm$0.02  &  4.8 $\pm$0.03 &   5.56$\pm$0.07 &   6.18$\pm$0.08  \\
  12575529-6456536   &  5068$\pm$73   &  2.79$\pm$0.25   &  1.15$\pm$0.12 &  7.42$\pm$0.01   &  0.09 &  7.42$\pm$0.03 &   7.53$\pm$0.16  &  7.42$\pm$0.05 &   6.22$\pm$0.03  &  4.77$\pm$0.02 &   5.44$\pm$0.07 &   6.11$\pm$0.05   \\
  12580262-6456492   &  4926$\pm$77   &  2.57$\pm$0.08   &  1.53$\pm$0.14 &  7.56$\pm$0.02  &  0.06 &  7.55$\pm$0.04 &   7.69$\pm$0.17  &  7.62$\pm$0.04 &   6.37$\pm$0.05  &  4.91$\pm$0.02 &   5.6 $\pm$0.05 &   6.23$\pm$0.01  \\
\hline		   											        
  18503724-0614364    &  4820$\pm$71   &  2.42$\pm$0.21   &  1.82$\pm$0.13 &  7.47$\pm$0.02  &  0.14 &  7.54$\pm$0.08 &   7.78$\pm$0.09  &  7.62$\pm$0.05 &   6.3 $\pm$0.07  &  4.83$\pm$0.0  &   5.62$\pm$0.04 &   6.21$\pm$0.07   \\
  18504737-0617184    &  4325$\pm$130  &  1.72$\pm$0.29   &  1.56$\pm$0.16 &  7.54$\pm$0.04  &  0.15 &  7.54$\pm$0.16 &   7.85$\pm$0.03  &  7.63$\pm$0.08 &   6.35$\pm$0.05  &  4.87$\pm$0.05 &   5.64$\pm$0.04 &   6.3 $\pm$0.01  \\
  18505494-0616182    &  4689$\pm$109  &  2.37$\pm$0.43   &  1.46$\pm$0.12 &  7.58$\pm$0.01  &  0.09 &  7.6 $\pm$0.1  &   7.8 $\pm$0.03  &  7.71$\pm$0.03 &   6.42$\pm$0.04  &  5.01$\pm$0.06 &   5.75$\pm$0.09 &   6.38$\pm$0.02   \\
  18505581-0618148    &  4577$\pm$139  &  2.23$\pm$0.31   &  1.60$\pm$0.24 &  7.69$\pm$0.03  &  0.18 &  7.61$\pm$0.08 &   7.89$\pm$0.08  &  7.8 $\pm$0.05 &   6.49$\pm$0.02  &  5.16$\pm$0.04 &   5.8 $\pm$0.09 &   6.49$\pm$0.01  \\
  18505755-0613461    &  4873$\pm$114  &  2.37$\pm$0.32   &  1.33$\pm$0.19 &  7.53$\pm$0.04  &  0.14 &  7.38$\pm$0.01 &   7.71$\pm$0.04  &  7.66$\pm$0.04 &   6.46$\pm$0.1   &  4.87$\pm$0.07 &   5.59$\pm$0.03 &   6.21$\pm$0.15   \\
  18505944-0612435   &  4925$\pm$177  &  2.56$\pm$0.39   &  1.50$\pm$0.5  &  7.66$\pm$0.05  &  0.18 &  7.68$\pm$0.07 &   7.88$\pm$0.12  &  7.68$\pm$0.06 &   6.42$\pm$0.12  &  5.01$\pm$0.04 &   5.79$\pm$0.07 &   6.4 $\pm$0.05  \\
  18510023-0616594    &  4433$\pm$95   &  1.94$\pm$0.47   &  1.50$\pm$0.14 &  7.63$\pm$0.02  &  0.12 &  7.58$\pm$0.13 &   7.91$\pm$0.1   &  7.71$\pm$0.13 &   6.42$\pm$0.02  &  5.05$\pm$0.02 &   5.75$\pm$0.07 &   6.45$\pm$0.01   \\
  18510032-0617183    &  4850$\pm$100  &  2.38$\pm$0.21   &  1.60$\pm$0.33 &  7.55$\pm$0.03  &  0.15 &  7.55$\pm$0.09 &   7.84$\pm$0.1   &  7.63$\pm$0.06 &   6.43$\pm$0.08  &  4.94$\pm$0.03 &   5.68$\pm$0.02 &   6.31$\pm$0.04  \\
  18510200-0617265   &  4415$\pm$87   &  2.35$\pm$0.45   &  1.48$\pm$0.07 &  7.7 $\pm$0.02  &  0.14 &  7.82$\pm$0.1  &   7.94$\pm$0.08  &  7.78$\pm$0.06 &   6.48$\pm$0.01  &  5.03$\pm$0.02 &   5.7 $\pm$0.02 &   6.47$\pm$0.02   \\
  18510289-0615301    &  4750$\pm$112  &  2.40$\pm$0.28   &  1.45$\pm$0.13 &  7.55$\pm$0.01  &  0.07 &  7.53$\pm$0.08 &   7.76$\pm$0.11  &  7.66$\pm$0.06 &   6.44$\pm$0.06  &  5.02$\pm$0.02 &   5.73$\pm$0.06 &   6.32$\pm$0.04  \\
  18510341-0616202    &  4975$\pm$146  &  2.50$\pm$0.3    &  1.94$\pm$0.27 &  7.59$\pm$0.07  &  0.15 &  7.46$\pm$0.07 &   7.97$\pm$0.06  &  7.58$\pm$0.07 &   6.42$\pm$0.07  &  5.15$\pm$0.02 &   5.75$\pm$0.06 &   6.31$\pm$0.0    \\
  18510358-0616112    &  4832$\pm$79   &  2.31$\pm$0.31   &  1.62$\pm$0.19 &  7.59$\pm$0.01  &  0.08 &  7.62$\pm$0.11 &   7.79$\pm$0.06  &  7.72$\pm$0.08 &   6.44$\pm$0.01  &  4.98$\pm$0.05 &   5.68$\pm$0.02 &   6.36$\pm$0.02  \\
  18510786-0617119   &  4768$\pm$53   &  2.11$\pm$0.19   &  1.80$\pm$0.28 &  7.53$\pm$0.01  &  0.14 &  7.49$\pm$0.04 &   8.03$\pm$0.01  &  7.67$\pm$0.11 &   6.45$\pm$0.02  &  5.0 $\pm$0.01 &   5.72$\pm$0.06 &   6.29$\pm$0.03   \\
  18510833-0616532    &  4750$\pm$112  &  2.25$\pm$0.22   &  1.60$\pm$0.25 &  7.66$\pm$0.05  &  0.10 &  7.62$\pm$0.04 &   7.93$\pm$0.1   &  7.66$\pm$0.07 &   6.45$\pm$0.02  &  4.96$\pm$0.02 &   5.78$\pm$0.12 &   6.44$\pm$0.02  \\
  18511013-0615486   &  4439$\pm$59   &  1.87$\pm$0.53   &  1.50$\pm$0.1  &  7.53$\pm$0.02  &  0.12 &  7.51$\pm$0.09 &   7.84$\pm$0.1   &  7.66$\pm$0.08 &   6.35$\pm$0.01  &  4.82$\pm$0.05 &   5.6 $\pm$0.03 &   6.34$\pm$0.04   \\
  18511048-0615470    &  4744$\pm$122  &  2.12$\pm$0.33   &  1.70$\pm$0.3  &  7.55$\pm$0.05  &  0.14 &  7.53$\pm$0.11 &   8.02$\pm$0.03  &  7.58$\pm$0.18 &   6.41$\pm$0.06  &  4.98$\pm$0.04 &   5.71$\pm$0.01 &   6.31$\pm$0.01  \\
  18511452-0616551    &  4800$\pm$59   &  2.40$\pm$0.25   &  1.69$\pm$0.2  &  7.6 $\pm$0.08  &  0.08 &  7.63$\pm$0.04 &   7.83$\pm$0.28  &  7.72$\pm$0.08 &   6.48$\pm$0.07  &  5.03$\pm$0.04 &   5.72$\pm$0.03 &   6.39$\pm$0.02   \\
  18511534-0618359    &  4755$\pm$57   &  2.16$\pm$0.21   &  1.79$\pm$0.17 &  7.58$\pm$0.02  &  0.22 &  7.53$\pm$0.01 &   7.99$\pm$0.09  &  7.72$\pm$0.1  &   6.45$\pm$0.03  &  5.03$\pm$0.04 &   5.77$\pm$0.04 &   6.36$\pm$0.03  \\
  18511571-0618146    &  4710$\pm$159  &  2.27$\pm$0.3    &  1.60$\pm$0.18 &  7.64$\pm$0.03  &  0.12 &  7.66$\pm$0.04 &   7.81$\pm$0.12  &  7.7 $\pm$0.04 &   6.45$\pm$0.04  &  5.0 $\pm$0.02 &   5.74$\pm$0.04 &   6.43$\pm$0.02   \\
  18512662-0614537   &  4459$\pm$91   &  2.10$\pm$0.48   &  1.48$\pm$0.19 &  7.61$\pm$0.01  &  0.17 &  7.61$\pm$0.09 &   7.93$\pm$0.14  &  7.8 $\pm$0.14 &   6.45$\pm$0.02  &  5.11$\pm$0.06 &   5.75$\pm$0.06 &   6.44$\pm$0.01  \\
  18514130-0620125    &  4671$\pm$140  &  2.20$\pm$0.28   &  1.62$\pm$0.2  &  7.61$\pm$0.03  &  0.19 &  7.55$\pm$0.09 &   7.8 $\pm$0.1   &  7.68$\pm$0.06 &   6.41$\pm$0.0   &  4.93$\pm$0.01 &   5.74$\pm$0.04 &   6.37$\pm$0.01\\
\hline \hline
\end{tabular}
\label{tab_dr1}\\
\end{table}
\end{landscape}

\begin{figure}
\centering
\includegraphics[width=8.5cm]{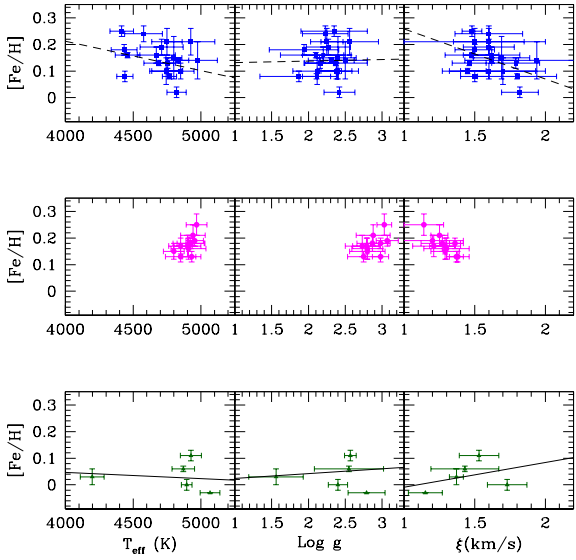}
\caption{[Fe/H] versus stellar parameters in the three clusters. 
The results of NGC~6705 (blue) are shown in the upper panels, while Trumpler~20 (magenta) is in the middle panels, and NGC~4815 (dark green) in the bottom panels. 
 }
\label{fig_par_feh}
\end{figure}

\begin{figure}
\centering
\subfigure[Temperature]
   {\includegraphics[width=7.5cm]{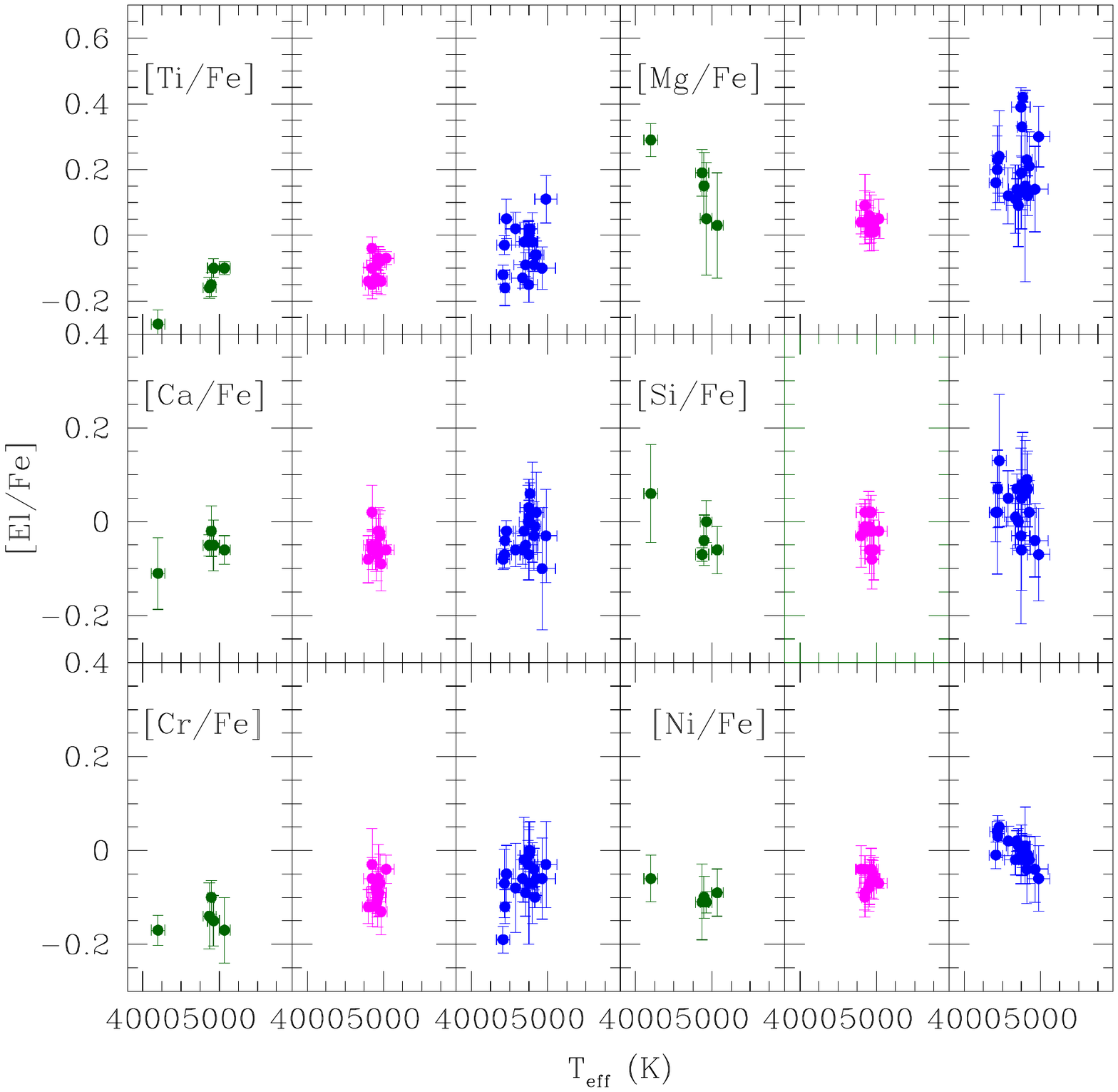}}
 \hspace{5mm}
 \subfigure[Gravity]
   {\includegraphics[width=7.5cm]{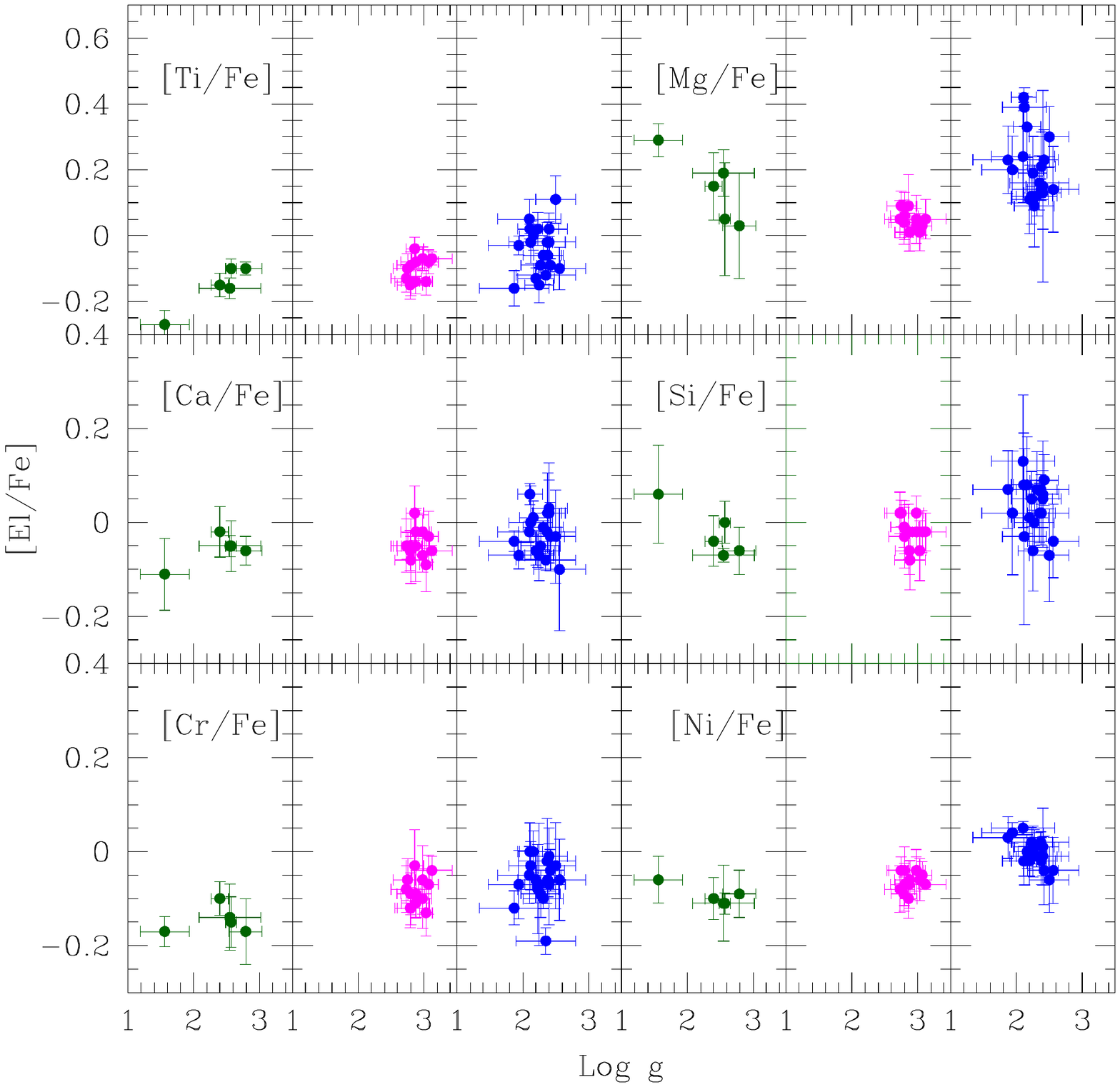}} 
   \hspace{5mm}
 \subfigure[Microturbolence]
   {\includegraphics[width=7.5cm]{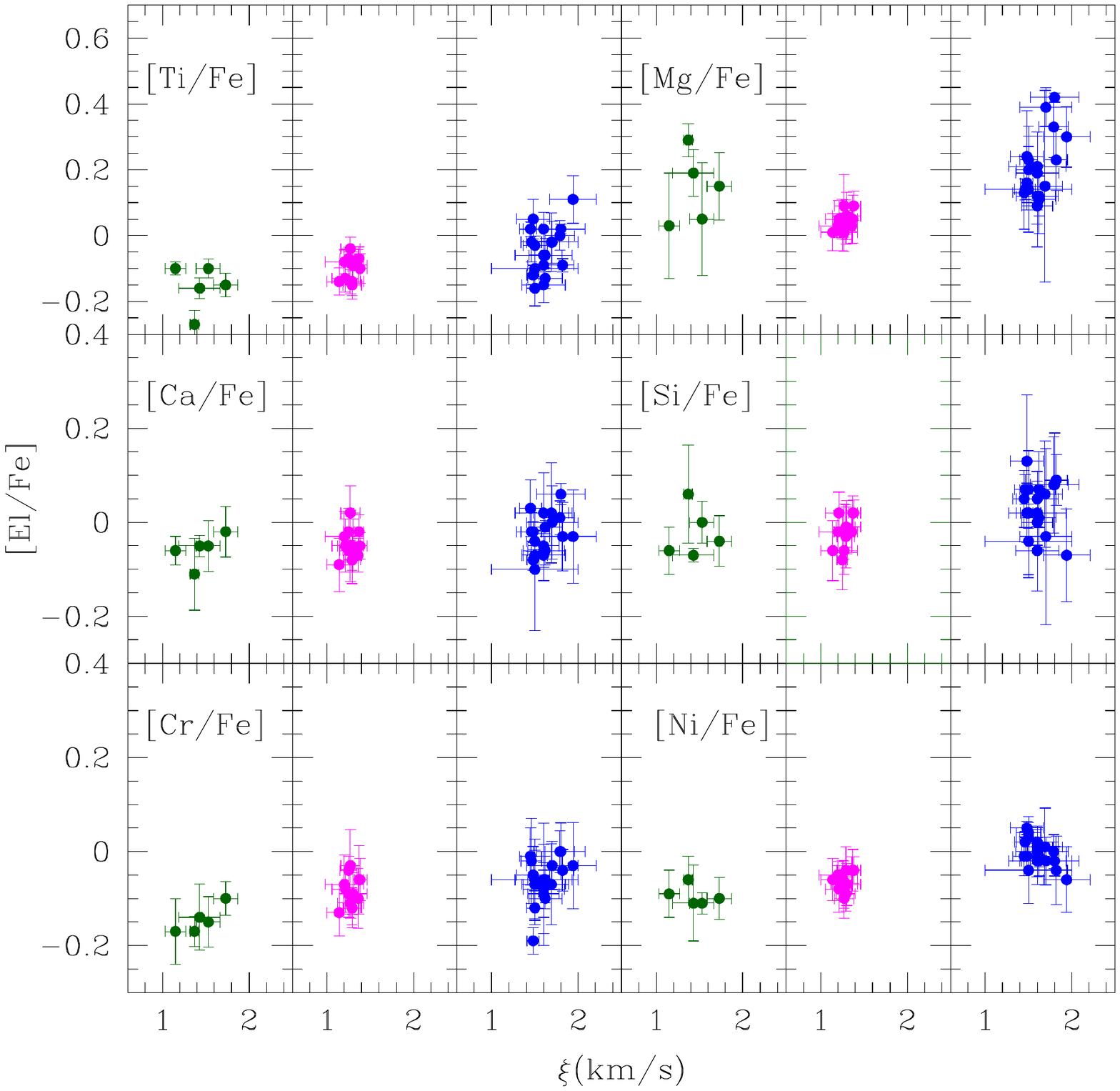}} 
   \hspace{5mm}
   \caption{Abundance ratios versus stellar parameters (panel (a) effective temperature, panel (b) gravity, panel (c) microturbolent velocity) in the three clusters. In each panel, the stars of NGC~4815 are in green, those of Trumpler~20 in magenta, and 
   those of NGC~6705 in blue.   }
\label{fig_par_el}
\end{figure}

The results are shown in Fig.~\ref{fig_par_feh} for [Fe/H] and in Figs.~\ref{fig_par_el} for [El/Fe]. 
In Fig.~\ref{fig_par_feh} we plot the stellar parameters versus [Fe/H] for member stars in the three clusters. We show also the 
mean least squares fits to the data for NGC~6705 and NGC~4815. We do not plot the linear fits for Trumpler~20, since they are artificially driven by the small 
interval spanned in stellar parameters by its member stars.  The trends 
are almost absent  for NGC~4815, while 
some trends are present for NGC~6705, even if the ranges in T$_{eff}$, log~g and $\xi$ spanned by its members are again quite small. 

In Figs.~\ref{fig_par_el}  the plots of [El/Fe] versus stellar parameters are shown. 
The trends with T$_{eff}$ and log~g are almost zero for all elements for NGC~6705 and Trumpler~20 where we have greater statistics, with a possible exception of [Mg/Fe]. 
The trends with $\xi$ seem to be more important, and might affect more elements. 
In general the cluster stars do not show important trends which could affect our further analysis. 

\subsection{Confirming the chemical homogeneity of clusters}
It is often stated that open clusters  are among the best objects  for tracing the star-formation history in the
disk  \citep[see, e.g.,][]{friel95}. They are considered to be composed of simple stellar populations, i.e., homogeneous  in terms of 
age and chemical composition. 
Thus, we investigate  the degree of homogeneity of the abundances of cluster members. 
We compare the standard deviation ($\sigma$) of all elements  with the average uncertainty in the abundances ($\Delta=<\delta>$), that is computed averaging the error on abundance ratios of each member star ($\delta$). 
We considered the cluster homogeneous in a specified element if 
the intrinsic scatter, as given by the $\sigma$, is lower or comparable to the average error $\Delta$, which should be indicative of the expected dispersion. 

\begin{figure}
 \centering
\includegraphics[width=8.5cm]{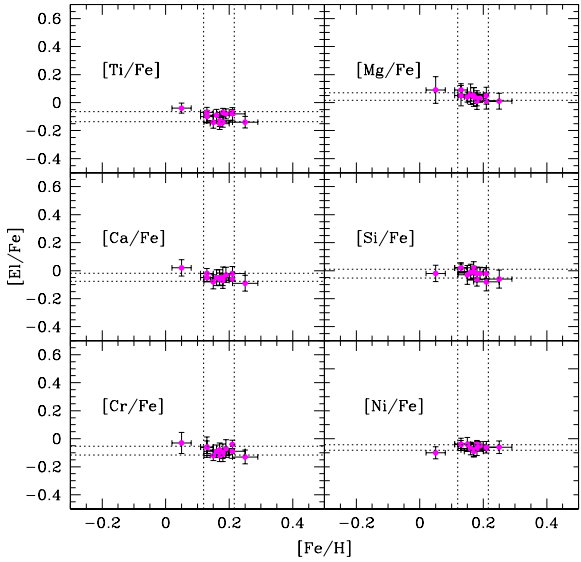}
 \caption{Abundance ratios versus [Fe/H] for individual member stars in Tr20. Errors on abundance ratios [El/Fe] are computed summing in quadrature 
 the errors in [El/H] and the errors on [Fe/H]. In each panel, the rectangular region shown by intersection of the four dotted lines indicate the 1-$\sigma$ area around the average value.    }
\label{fig_tr20}
 \end{figure}

\begin{figure}
 \centering
\includegraphics[width=8.5cm]{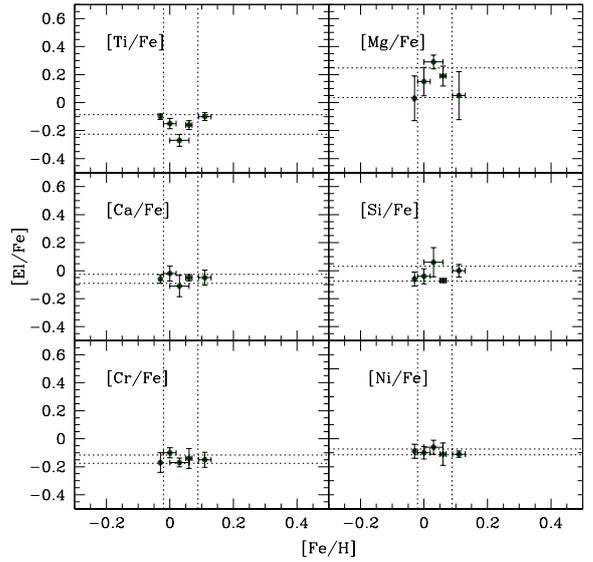}
 \caption{As in Fig.~\ref{fig_tr20}, but for NGC~4815.  }
\label{fig_ngc4815}
 \end{figure}

\begin{figure}
 \centering
\includegraphics[width=8.5cm]{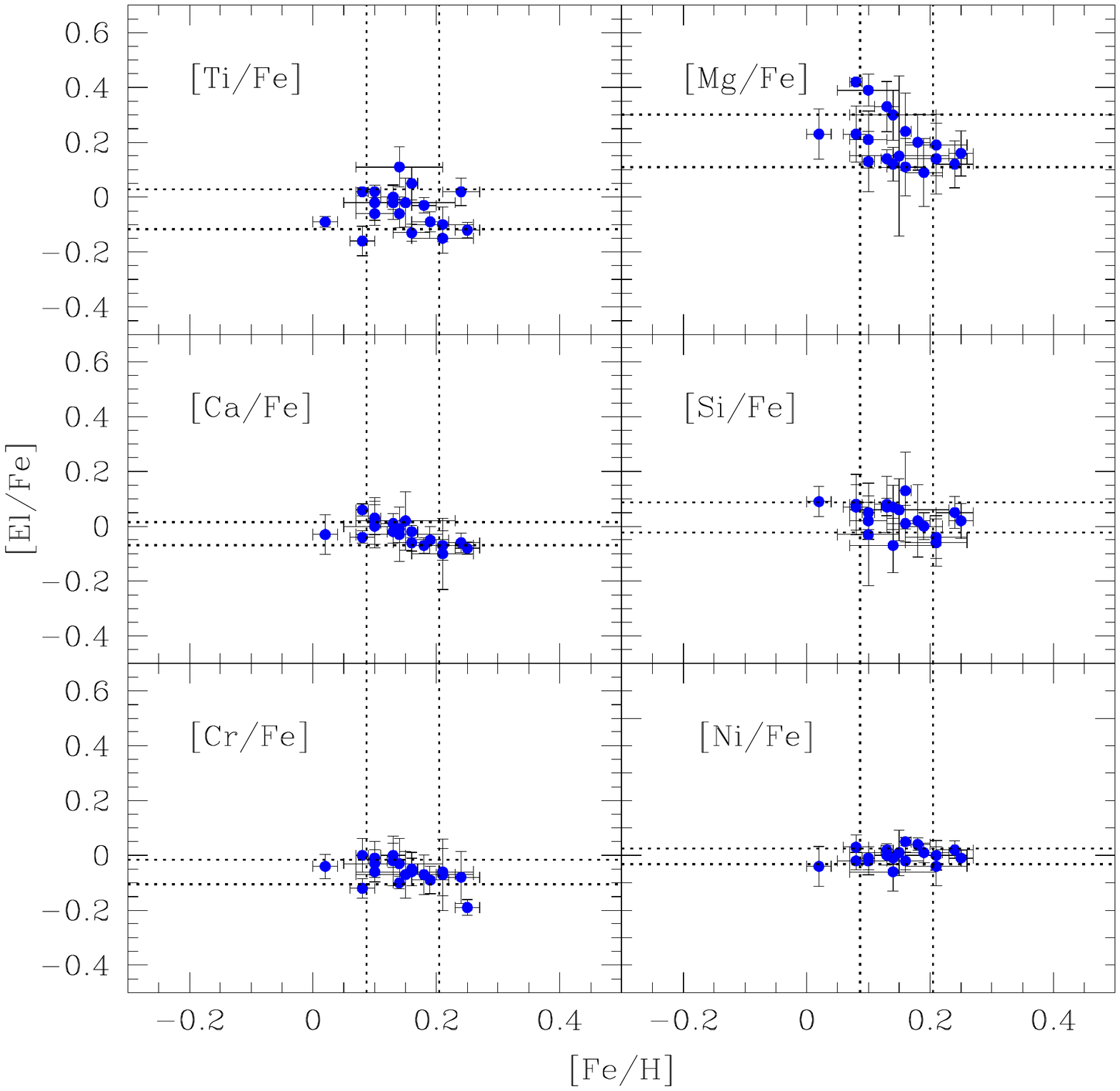}
 \caption{As in Fig.~\ref{fig_tr20}, but for NGC~6705.   }
\label{fig_ngc6705}
 \end{figure}

The results are shown in Table~\ref{tab_scatter} and in Figs.~\ref{fig_tr20}, \ref{fig_ngc4815}, \ref{fig_ngc6705}. 
The rectangles indicate the regions 1-$\sigma$ wide around the average. 
The three clusters are essentially homogeneous in all elements. 
For Trumpler~20 we notice the presence of a star ($\#$12391577-6034406) that is slightly metal poorer than the main body of  cluster members. 
However, its radial velocity and stellar parameters are in agreement with those of the  other 
observed clump stars.  
As discussed in Donati et al. (2013), the Besan\c con model \citep{robin03} computed at the location of Trumpler~20 gives that 
indicatively 17\% of the candidate members for radial velocities may be still field stars. This corresponds to $\sim$1-2 stars being 
possibly non-members in a sample of 13 stars, and it could justify the lower [Fe/H] of  $\#$12391577-6034406.
For NGC~6705 we note a possible bi-modal behaviour of [Mg/Fe] with four stars outside the 1-$\sigma$ rectangle around the average (see also the trends in  Figs.~\ref{fig_par_el}).

In Table~\ref{tab_scatter} we report for each cluster:  the abundance ratios,  the standard deviation $\sigma$  and the average error on each measurement, $\Delta$, and the number of stars used to compute these quantities. For [El/Fe] ratios $\Delta$ take into account the errors on [El/H] and on [Fe/H], summed in quadrature. 
We note, however, that this approach might overestimate the error on [El/Fe] since  the error on [El/H] and [Fe/H] are likely correlated. 
The best approach should be to consider the effect on [El/Fe] due to the errors on  the atmospheric parameters, but this kind of errors are not included in the present release.
For the error on [Fe/H] we used the values in column $\sigma$(Fe~I) of Table~2.
If we consider the best determined element, i.e. iron, 
we note that $\sigma$  is lower than  $\Delta$. 
This is true for most of the other elements, with the possible exception of Mg and Ti which have a larger scatter in NGC~6705 and NGC~4815. 
Thus we can conclude that, with the present level of precision, these open clusters are homogeneous with respect  to their content in $\alpha$-elements (Ti, Si, Ca, Mg) and iron-peak elements (Fe, Ni, Cr).  
In Figs.~\ref{fig_tr20}, \ref{fig_ngc4815}, \ref{fig_ngc6705} these results are reinforced: the abundance ratios of clusters do not show any correlation with [Fe/H], and within 
the errors, they are homogeneous.

 \begin{table}
\begin{center}
\caption{$\sigma$ and  $\Delta$ for abundance ratios in each cluster. }
\tiny
\begin{tabular}{lrll}
\hline\hline
\hline
El           & Mean        &     $\sigma$$^{(a)}$  &   $\Delta$$^{(b)}$  \\
\hline \hline
Tr20 & & & \\
\multicolumn{4}{l}{N (number of member stars)=13}\\
\hline
$[$Fe/H$]$   &      0.17 &            0.05 &        0.07  \\
$[$Si/Fe$]$  &     -0.02 &            0.03 &        0.08      \\
$[$Ca/Fe$]$  &      -0.05 &            0.03 &        0.08    \\
$[$Mg/Fe$]$  &      0.04 &            0.03 &        0.09     \\
$[$Ti/Fe$]^{(c)}$  &     -0.10 &            0.04 &        0.08   \\
$[$Ni/Fe$]$  &     -0.06 &            0.05 &        0.08     \\
$[$Cr/Fe$]$  &     -0.08 &            0.02 &        0.08   \\
\hline
NGC~4815  & & & \\
\multicolumn{4}{l}{N (number of member stars)=5}\\
\hline
$[$Fe/H$]$  &     0.03  &            0.05 &       0.10    \\
$[$Si/Fe$]$ &     -0.02   &            0.05 &       0.09    \\
$[$Ca/Fe$]$ &    -0.06   &            0.03 &       0.09     \\
$[$Mg/Fe$]$ &     0.14   &            0.10 &       0.13    \\
$[$Ti/Fe$]^{(a)}$ &    -0.16   &            0.07 &       0.08  \\
$[$Ni/Fe$]$ &    -0.09   &            0.02 &       0.09    \\
$[$Cr/Fe$]$ &    -0.15   &            0.03 &       0.09   \\
\hline
NGC~6705 & && \\
\multicolumn{4}{l}{N (number of member stars)=21}\\
\hline
$[$Fe/H$]$   &    0.14  &             0.06      &   0.14    \\ 
$[$Si/Fe$]$  &    0.03  &             0.05       &  0.16    \\
$[$Ca/Fe$]$  &    -0.02  &             0.05     &    0.14    \\
$[$Mg/Fe$]$  &    0.20  &             0.09      &   0.17  \\
$[$Ti/Fe$]^{(a)}$  &   -0.05  &             0.07        & 0.14   \\
$[$Ni/Fe$]$  &    0.01  &             0.03        & 0.16    \\
$[$Cr/Fe$]$  &   -0.07  &             0.05       &  0.15    \\
\hline \hline
\end{tabular}
\label{tab_scatter}\\
\end{center}
(a) $\sigma$: standard deviation of all elements; (b) $\Delta$: the average uncertainty in the abundances ($\Delta=<\delta>$)\\
(c) [Ti/Fe] is computed using only Ti~I lines. 
\end{table}


 \subsection{Chemical patterns}
\begin{figure}
 \centering
\includegraphics[width=8.5cm]{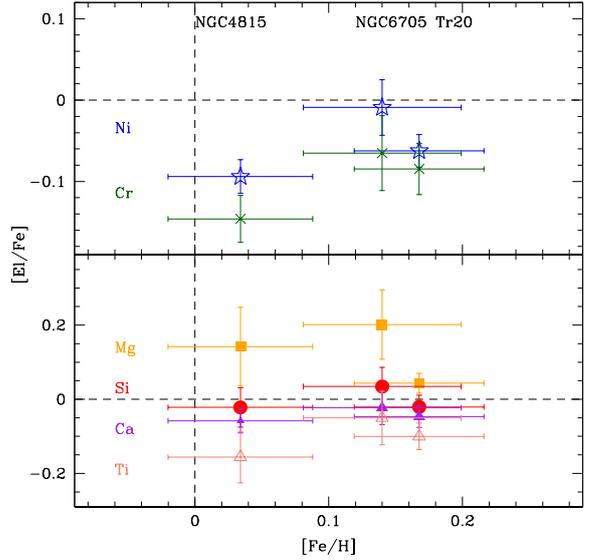}
 \caption{Average values of the four $\alpha$-elements  (Ca, Si, Ti, Mg) in the bottom panel. Results of iron-peak elements (Ni, Cr) 
are shown in the upper panels. Clusters are ordered by [Fe/H], NGC 4815 is the first cluster on the left, NGC 6705 in the middle, and Trumpler 20 on the right side. 
The errors  are the  standard deviation,  $\sigma$, computed  for member stars of each clusters. The colour and symbol code is the following: green crosses for [Cr/Fe], 
 blue stars for [Ni/Fe],  red circles for [Si/Fe], purple filled triangles for [Ca/Fe], orange squares for [Mg/Fe], and  empty salmon triangles for [Ti/Fe].  }
\label{fig_scatter}
 \end{figure}

Having established the chemical homogeneity of these elements in the three clusters under analysis, 
we can compare their abundance patterns using the average abundances as representative of the entire  cluster. 
In Fig.~\ref{fig_scatter}, we present graphically these results. 
An important feature  is the comparison of the average abundance ratios in the three open clusters. 
As said in the previous sections, Trumpler~20, NGC4815, and NGC~6705 are located at similar distances from the Galactic Centre. They mainly differ 
in terms of age, cluster mass, and  metallicity [Fe/H].
In terms of [Fe/H], the metal poorest cluster (NGC~4815) and the metal richest one (Trumpler~20) differ $\sim$2-$\sigma$. 

\begin{figure}
 \centering
\includegraphics[width=8.5cm]{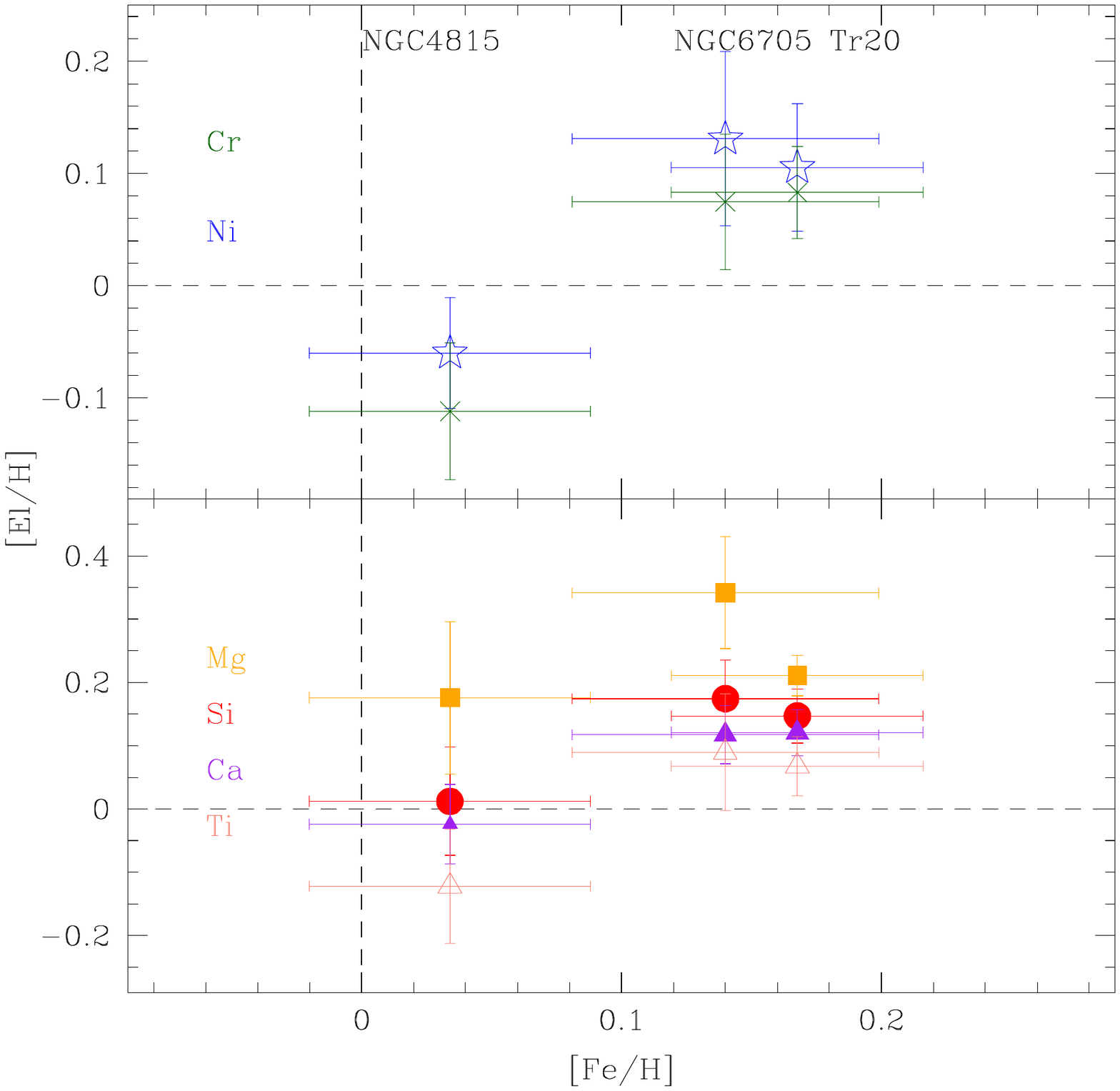}
 \caption{As in Fig.\ref{fig_scatter}, but for [El/H].  }
\label{fig_scatter_elh}
 \end{figure}

Inspecting Fig.~\ref{fig_scatter}, we note that each cluster shows unique features with respect to the other clusters: 
Trumpler~20 has  solar [Si/Fe], [Mg/Fe] and [Ca/Fe], slightly depleted in [Ti/Fe], [Ni/Fe] and [Cr/Fe]. 
NGC~4815 is solar in [Si/Fe] and [Ca/Fe] and slightly depleted [Ti/Fe], [Cr/Fe] and 
[Ni/Fe], and enhanced in [Mg/Fe].  
NGC~6705 has solar [Ti/Fe], [Si/Fe], [Ca/Fe] and [Ni/Fe], while it is enhanced in [Mg/Fe] and depleted in [Cr/Fe].   
We statistically quantified the level of significance of the cluster-to-cluster abundance differences estimating, for each abundance ratio in each 
pair of clusters, the following quantity 
\begin{equation}
(\rm {[El/Fe]_{cluster1} - [El/Fe]_{cluster2}})/ \sqrt(\delta(\rm {[El/Fe]_{cluster1}})^2+\delta(\rm {[El/Fe]_{cluster2}})^2),  
\end{equation}
where the abundance ratios and their errors are those reported  in Table~\ref{tab_scatter}. 
All pairs of comparison are of the order of 1-$\sigma$, with the  lowest difference for [Si/Fe] in Trumpler~20 and NGC~4815, 
and the largest difference for [Ni/Fe] in NGC~6705 and NGC~4815 that differ more than 2-$\sigma$.
This suggests that, in terms of abundance ratios over iron, these clusters are statistically very similar, but they differ in their global content of metals.
We have thus done the same comparison with the elemental abundances, [El/H]. The results are shown in Fig.\ref{fig_scatter_elh}, from which we see 
how the abundances of NGC~4815 differ from those of Trumpler~20 and NGC~6705 for all elements. 
We recomputed the level of significance of the cluster-to-cluster abundance differences finding that 
NGC~4815 differs $\sim$2-$\sigma$ from the other two clusters in most of its [El/H] abundances.   

These differences, especially those of [El/H],  can be considered intrinsic characteristics of the chemical composition of the interstellar medium (ISM) from which each cluster was born
since the analysis was performed  in a fully homogeneous way, from the target selection, to the observational strategy,  data reduction, and abundance analysis. 
A tentative first conclusion might be that, even if at present Trumpler~20, NGC~4815 and NGC~6705 are located at similar distances from the Galactic Centre (GC) (although NGC 6705 is on the opposite side of the Sun-Galactic Centre line), they did not originate from an ISM with the same composition. 
In particular the difference in their mean metallicity is  relevant, NGC~4815 having a lower metallicity (within 2-$\sigma$) than the other two clusters.
This can be obtained,  at least,  in three different ways: 
{\em i)} the ISM is not azimuthally homogeneous, and areas located at similar radii might have a different chemical composition due to local enrichment 
and to incomplete mixing; 
{\em ii)} the clusters might have moved from their place of birth, and thus they might reflect the chemical composition at a different radius with respect to their present  
position; 
{\em iii)} due to the different ages of the clusters, their mean metallicity and abundance ratios might be a signature of the temporal chemical  evolution of the Galactic disk. 
A comparison with the field population,  in the approximation that the migration does not dominate its distribution,  is useful to check these hypotheses and 
in Sec.\ref{sec_model} we describe a comparison with two different Galactic chemical evolution models to search for a possible  explanation about  the origin
of the abundance ratios in these inner-disk open clusters.

\subsection{A comparison of abundance distributions in the field and clusters}

The uniformity of the analysis of cluster and field stars offers the possibility to see differences in abundances at a level not possible before, 
when heterogeneous samples and analysis methods were considered.
We remind that our comparison is done with stars belonging to  different evolutionary stages, such as solar neighbourhood turn-off stars, open
cluster clump giants, and inner disk giants.  Even though these samples are
subjected to a homogeneous analysis, systematic offsets in abundances could arise.
Indeed, as a matter of caution when comparing stars in different evolutionary stages, we shall recall 
a recent analysis of stellar spectra for stars in the open cluster M 67. 
\citet{onehag11} analysed a solar twin in M 67,  a dwarf star with stellar
parameters very similar to those of the Sun, and found an [Fe/H] ratio
of $+$0.02 dex and [El/Fe] within 0.03 dex of the solar values. This is in contrast
to analysis of evolved giant stars in the same cluster for which \citet{yong05}  found an [Fe/H] ratio very close to the solar value but with abundance ratios that differ significantly from the solar
values. 
However, we consider that, more than the difference in evolutionary stages, different methods and different atomic data 
might affect the results of the analysis of \citet{onehag11} and of  \citet{yong05}. 
A contrasting example can be found in the study of IC~4615 \citep{pasquini} who analysed both dwarf and giant stars, for which no differences in the chemical composition were found. 

As also pointed by \citet{melendez08}, the use of the same set of lines and the choice of a common solar abundance scale for the normalisation of the stellar results are of great importance, and are among the main strengths of our analysis.   
We recall that different conclusions were drawn for instance from a comparison of bulge and thick disk stars in \citet{fulbright07} and \citet{melendez08}. 
Those differences are probably due to the  heterogeneous comparison in the former paper of their bulge giant results with literature values for main sequence and turn-off disk stars in the solar neighbourhood \citep{bensby05, reddy06} analysed with a different line list and normalised to a different solar zero-point. 
In the Gaia-ESO survey analysis, all stars are analysed as homogeneously as possible, and these systematic effects 
should be reduced.

\begin{figure*}
\centering
\includegraphics[width=1\textwidth]{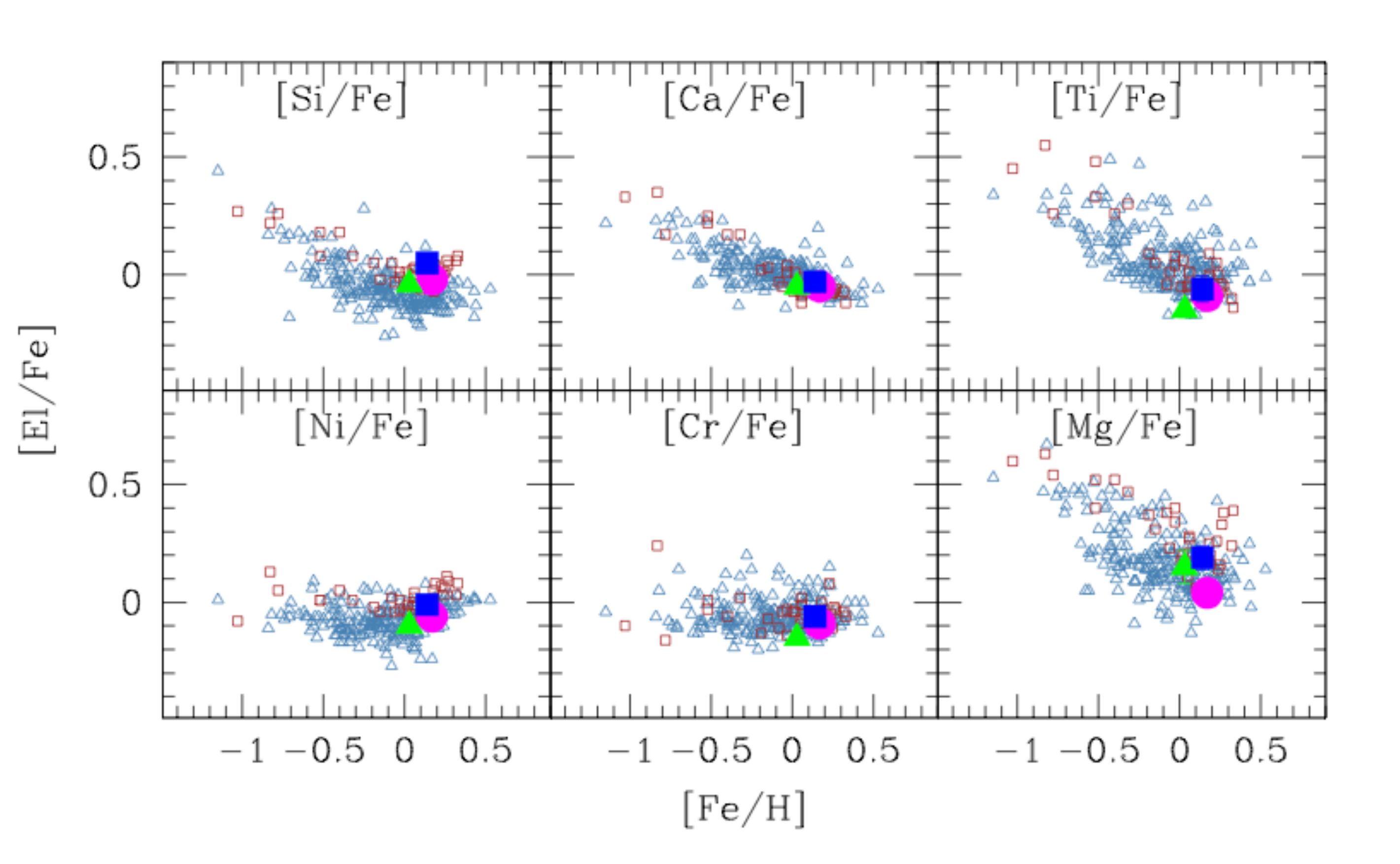}
\caption{Abundance ratios in field and open cluster stars: the abundance ratios [El/Fe] versus [Fe/H] of solar neighbourhood dwarf stars (cyan empty triangles), of inner disk/bulge giant stars (red empty squares). Clusters are represented by their average values:  Trumpler~20 (magenta filled circle), 
NGC~4815 (green filled triangle), and NGC~6705 (blue filled square). }
\label{Fig_field}
\end{figure*}

In Fig.~\ref{Fig_field} we show the abundance ratios [El/Fe] versus [Fe/H] of solar neighbourhood dwarf stars, of inner disk/bulge giant stars and of clusters. We note that elemental abundances of stars in  open clusters are consistent, within the errors, with the  trends of [El/Fe] versus [Fe/H] for almost all elements in field stars.  
However cluster stars show some differences from field stars having the same [Fe/H]. 

\begin{figure}
 \centering
 \includegraphics[width=9.cm]{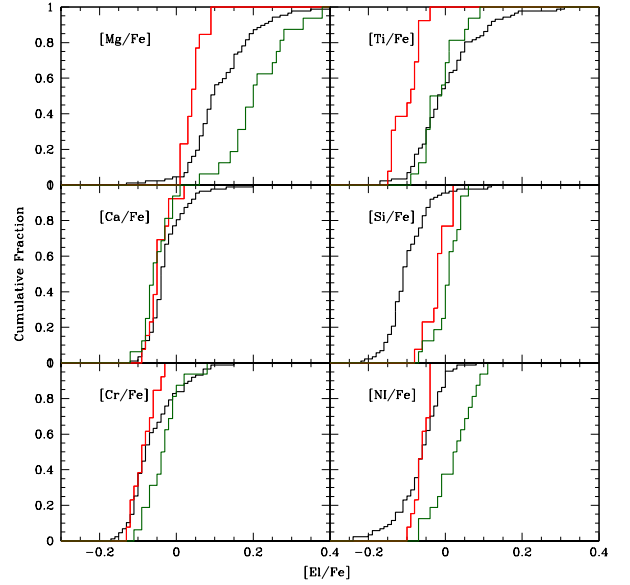}
 \caption{Comparison of cumulative distribution of the Trumpler~20 abundance ratios (red curves) with the solar neighbourhood turn-off stars with the same metallicity  (black curves) and with the inner disk/bulge giant stars (green curves).  }
\label{fig_ks_tr20}
 \end{figure}

\begin{figure}
 \centering
 \includegraphics[width=9.cm]{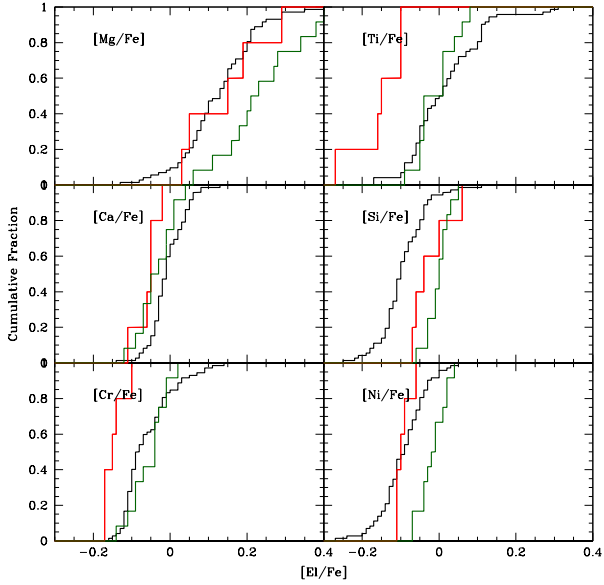}
 \caption{As in Figure~\ref{fig_ks_tr20}, but for NGC~4815.}
\label{fig_ks_4815}
 \end{figure}

\begin{figure}
 \centering
 \includegraphics[width=9.cm]{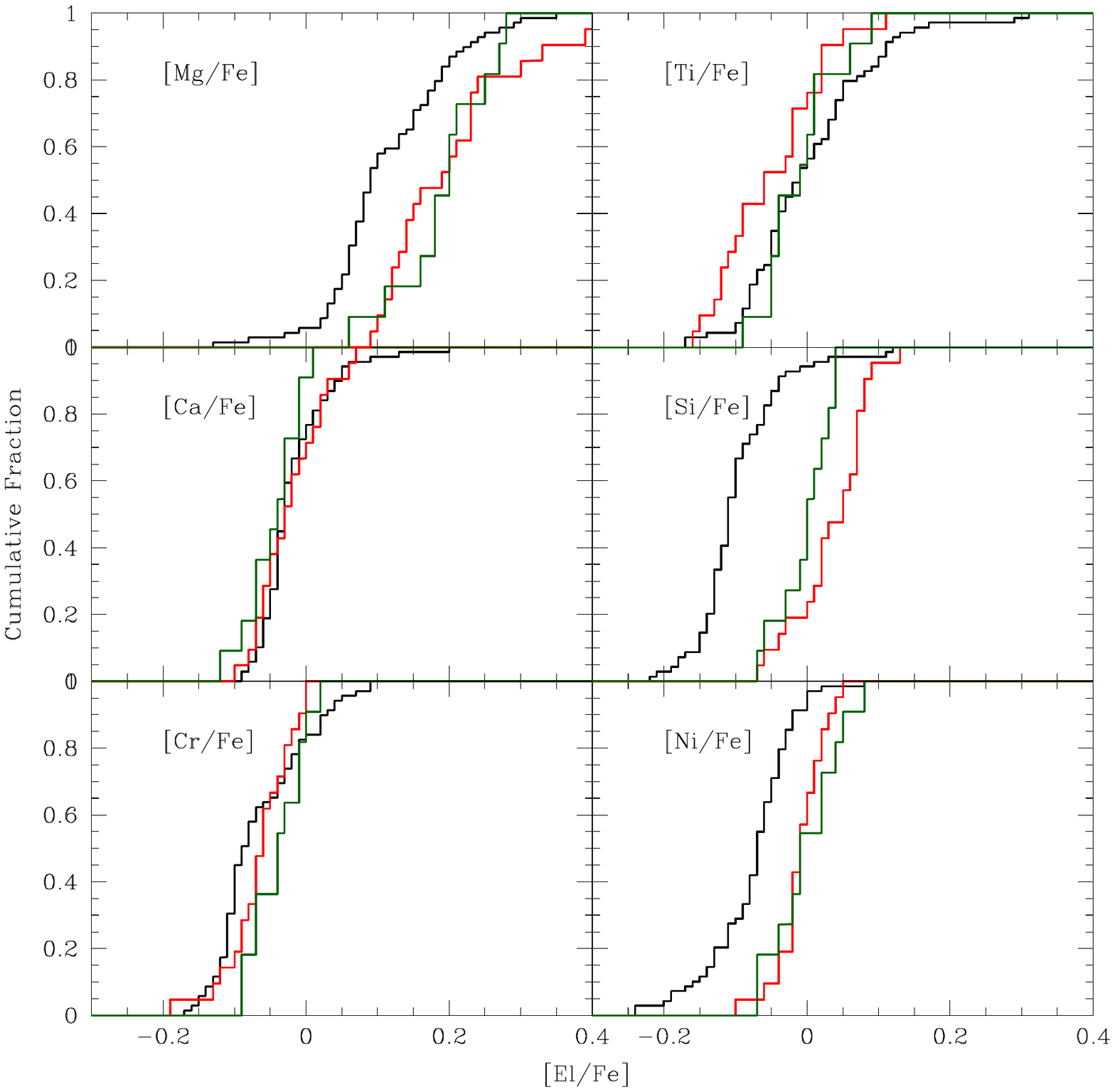}
 \caption{As in Figure~\ref{fig_ks_tr20}, but for NGC~6705.}
\label{fig_ks_6705}
 \end{figure}

A statistical way to compare two distinct populations is to compare their cumulative distributions. In our case we want to probe possible differences in the 
chemical composition of two populations, thus we compare the cumulative distribution  of their elemental abundance ratios. 
For each cluster we have selected for the comparison  only  field stars 
(both solar neighbourhood and inner-disk/bulge stars)  in the same metallicity range, 0.2 dex, centred around the mean 
metallicity of the cluster stars: 0.05$\leq$[Fe/H]$<$0.25 for Trumpler~20, 0.0$\leq$[Fe/H]$<$0.2 for  NGC6705 and   -0.1$\leq$[Fe/H]$<$0.1 for NGC~4815.  
The cumulative distributions of abundance ratios are shown in  Figs.~\ref{fig_ks_tr20}, \ref{fig_ks_4815}, \ref{fig_ks_6705}. 
The closer two distributions are, the higher is the probability that they come from populations sharing the same chemical composition. 
For instance, [Ca/H] distribution in Tr~20 and in the solar neighbourhood stars  have a probability of $\sim$60\% 
to derive from the same  population, while for [Mg/Fe] the probability is lower than 1\%. 
For NGC~4815, the highest probabilities of similar distributions are for [Mg/Fe] ($\sim$90\%), [Ni/Fe] ($\sim$50\%), [Si/Fe] ($\sim$20\%) 
in NGC~4815 and in the solar neighbourhood. 
For NGC~6705, the probabilities that the cluster and inner-disk stars came from similar populations 
are: [Cr/Fe] ($\sim$30\%), [Ca/Fe] ($\sim$50\%), [Mg/Fe] ($\sim$70\%), [Ni/Fe] ($\sim$80\%),
[Ti/Fe] ($\sim$10\%). 
Due to the small number  statistics of our  analysis we recall that the  probabilities associated with the statistical 
Kolmogorov-Smirnov test have a limited confidence and are only indicative.

In Fig.~\ref{fig_ks_tr20} we have the results for Trumpler~20: 
this cluster is indistinguishable from the field population in its iron-peak elements. 
The alpha-elements do not all behave  in the same way: two of them 
are remarkably under-abundant compared to field stars with the same metallicity (Mg and Ti), 
while Ca and Si are distributed similarly to the inner-disk/bulge stars.
In Fig.~\ref{fig_ks_4815} we show  the results for NGC~4815, which
exhibits a clear under-abundance of Ti, 
and has  [Si/Fe]  between the solar neighbourhood and inner disk stars, 
while Mg and Ca are similar to the solar neighbourhood sample.
Among the iron-peak elements Cr is  slightly lower than that in the solar neighbourhood sample, while Ni  has the same distribution. 
In Fig.~\ref{fig_ks_6705} we present the results for NGC~6705:  
the behaviour of this cluster is very similar to that of stars located 
in the inner-disk/bulge, having similar distribution of Mg, Ti, Ca, Cr, and Ni. 
Only Si is enhanced with respect to both solar neighbourhood and inner-disk stars.

\subsection{Ti abundances in stars of different type} 
\begin{figure*}
 \centering
\includegraphics[width=16cm]{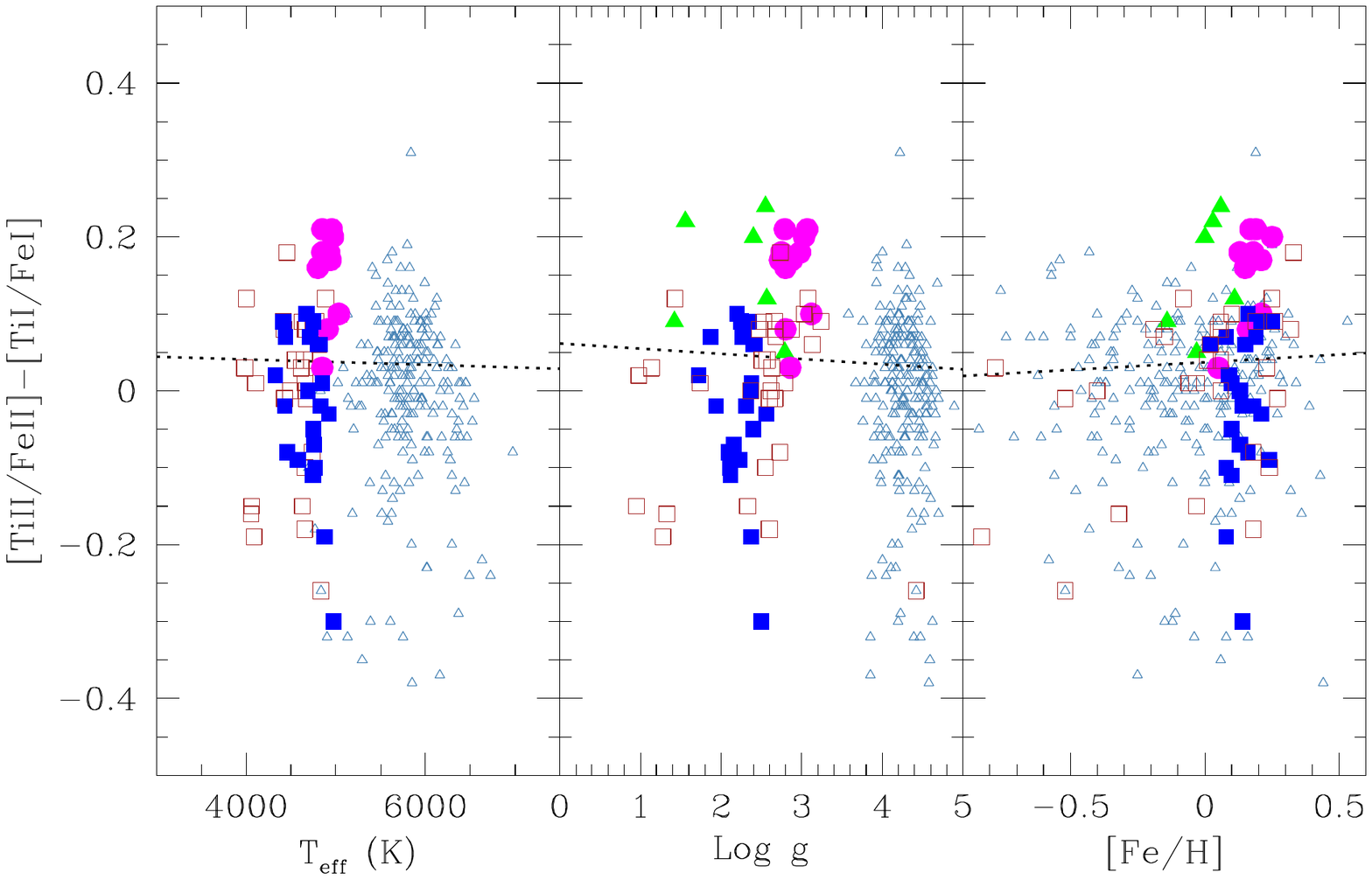}
 \caption{[TiII/Fe]-[TiI/Fe] vs. T$_{eff}$, log g, and [Fe/H] in the Milky Way field stars and in the cluster member stars.  Symbols as in Fig.~\ref{Fig_field}. The dotted lines are the weighted least mean square fits to the data. }
\label{fig_ti1ti2}
 \end{figure*}

Titanium is evidently more under-abundant in Trumpler~20 and NGC~4815 than in the field population with the same [Fe/H] (see Figs.~\ref{fig_ks_tr20} and \ref{fig_ks_4815}). 
Thus it might be a good tracer of the different abundance patterns of these two populations. 
To prove that this difference is real and not due to
NLTE over-ionisation effects in the considered range of stellar parameters, we have plotted   [TiII/FeII]-[TiI/FeI] vs. T$_{eff}$, as done by \citet{dorazi10}, and also vs. log~g and $\xi$. 
In their study of the open clusters IC~2602 and IC~2391, \citet{dorazi10} found indeed some trends of [El/Fe] vs. T$_{eff}$ for various elements. 
These effects were already found for old field stars by \citet{boda03} and by \citet{gilli06}, who explained them as due to NLTE effects. 
Other explanations were given from \citet{adibekyan12} who analysed a large sample of F-G-K stars for which they studied the relation 
between  [Ti/TiII]  and T$_{eff}$. They found a trend that they explained with problems associated with the differential analysis or with an incorrect opacity in the model atmospheres. 
For Ca and Ti,  \citet{dorazi10} obtained decreasing trends with temperature, with cooler stars having lower [Ca/Fe] and [Ti/Fe] than hotter stars in the same cluster. 
They proposed that the young age of their stars and the enhanced levels of chromospheric activity  affect them by NLTE over-ionisation (see their Fig.~4). 
This trend is completely absent in the three panels of Fig.~\ref{fig_ti1ti2}: [TiII/FeII]-[TiI/FeI] is almost consistent with zero along the whole T$_{eff}$, log~g and $\xi$ ranges. 
This is likely due to the older age of our stars, and reassures us about the use of titanium as a reliable tracer of the abundance patterns of open clusters.  
We did also a computation of the NLTE effects on a sample of TiI and TiII lines employed in the analysis of the Gaia-ESO Survey. 
We found that on average the NLTE correction for TiI are  of the order of $+$0.06~dex, while TiII is not affected by NLTE effects (Bergemann, private communication).

\section{On the origin of the abundance ratios in the inner open clusters}
\label{sec_model}

In this section we discuss what we can learn about the origin of the inner disk open clusters from their abundance patterns. 
We discuss only  elements  which are not affected by stellar evolution, and thus direct tracers of the ISM composition at the epoch when the cluster formed: the $\alpha$- and iron-peak elements. 
We compare their abundance ratios in  the field and cluster populations with the predictions of two different  chemical evolution models (Magrini et al. 2009, Romano et al. 2010, hereafter M09 and R10, respectively), briefly described below. 

\paragraph{Model of Romano et al. (2010)}

The model of R10 is based on the two-infall model case B for the chemical evolution of the Galaxy. 
A thorough discussion of the adopted formalism and basic equations can be found in \citet{chiappini97,chiappini01}. 
Here we briefly recall the overall evolutionary scenario.
The inner halo and thick disk of the Milky Way are assumed to form on a relatively short timescale (about 1~Gyr) out of a first infall episode, 
whereas the thin disk forms inside-out \citep{matteucci89} on longer timescales (7 Gyr in the solar vicinity) 
during a second independent episode of extragalactic gas infall. The Galactic disk is approximated by several independent rings, 2 kpc wide. 
Radial flows and outflows are not considered here.  
The adopted star formation rate (SFR) is proportional to both the total mass and the gas surface densities. The efficiency of conversion of gas into stars is higher during the halo/thick-disk phase than during the thin-disk phase. Furthermore, it drops to zero every time the gas density drops below a critical density threshold. 
The stellar lifetimes are taken into account in detail. As for the stellar IMF, the \citet{kroupa93} IMF is assumed in the 0.1--100 M$_\odot$ mass range. The rate of SNIa explosions is calculated as in 
\citet{matteucci86}. 
SNeIa explode in close binary systems when a CO white dwarf has reached a critical mass limit because of accretion of hydrogen-rich matter from a main-sequence or red giant companion.
The yields for SNeIa are taken from \citet{iwamoto99},  model W7. As for single stars, several sets of stellar yields are analysed by R10 (see their table~2). Here we show the results of their model~15, that adopts the yields by \citet{karakas10} for low- and intermediate-mass stars (1--6~M$_\odot$), the yields of rotating massive stars by the Geneva group (see R10, their table~1 for references) for helium, carbon, nitrogen and oxygen, and the yields of \citet{koba06} for heavier elements produced by Type II supernovae and hypernovae with progenitor masses in the range 13--40~M$_\odot$. The available stellar yields are interpolated in the mass range 6--13~M$_\odot$ and extrapolated up to 100~M$_\odot$ (see R10 for details). The tabulated yields are 
adopted as published, i.e. without  \emph{ad hoc} adjustments to reproduce the data. This notwithstanding, model~15 provides a good fit to the abundance ratios of most chemical elements in Milky Way's stars.

\paragraph{Model of Magrini et al. 2009}

The model adopted by M09 is a generalisation of the multi-phase model by \citet{ferrini92}, originally built for the solar neighbourhood, and subsequently extended to the entire Galaxy \citep{ferrini94} and to other disk galaxies \citep[e.g.,][]{molla96,molla05,magrini07}. 
The detailed description of the general model is in the above mentioned papers. 
The main assumptions of the model are that the Galaxy disk is formed by infall of gas from the halo and from the intergalactic medium. The adopted infall follows an exponentially decreasing law. This produces an inside-out formation scenario where the inner parts of the disk evolve more rapidly than the outer ones. 
As in the R10 model, radial flows and stellar migrations, as well as gas outflows are not considered.  
The stellar lifetimes are taken into account in details and for the stellar yields the following choices were made: 
for low- and intermediate-mass stars (M$<$8 $M_\odot$) they use the yields by \citet{gavilan05} for both values of the metallicity (Z=0.006 and Z=0.02). For stars in the mass range $8~M_\odot < M < 35~M_\odot$ they adopted the yields by \citet{chieffi02} for Z=0.006 and Z=0.02.
They estimated the yields of stars in the mass range $35~M_\odot <M<100~M_\odot$, which are not included in the tables of Chieffi \& Limongi (2004), by linear extrapolation of the yields in the mass range $8~M_\odot < M < 35~M_\odot$. In the case of Ti, M09 increased the yields by a factor of 2 to reproduce the observations. The rate of SNIa explosions is calculated as in \citet{matteucci86} and their  yields are taken from the model CDD1 by \citet{iwamoto99}. The IMF by \citet{kroupa93} is adopted.

\paragraph{Comparison models vs data}

\begin{figure*}
\centering
\subfigure[R10]
   {\includegraphics[width=10.5cm]{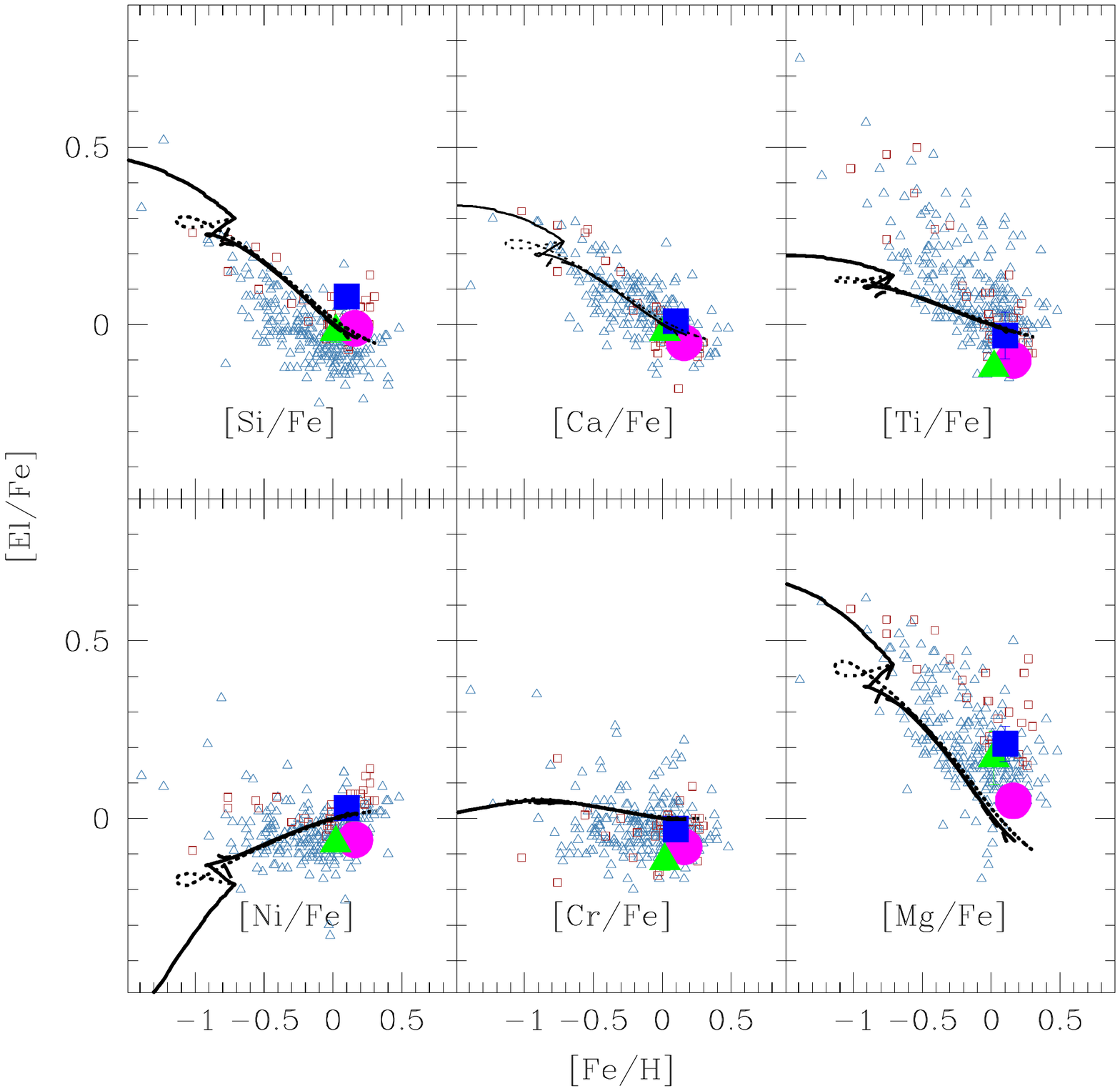}}
 \hspace{5mm}
 \subfigure[M09]
   {\includegraphics[width=10.5cm]{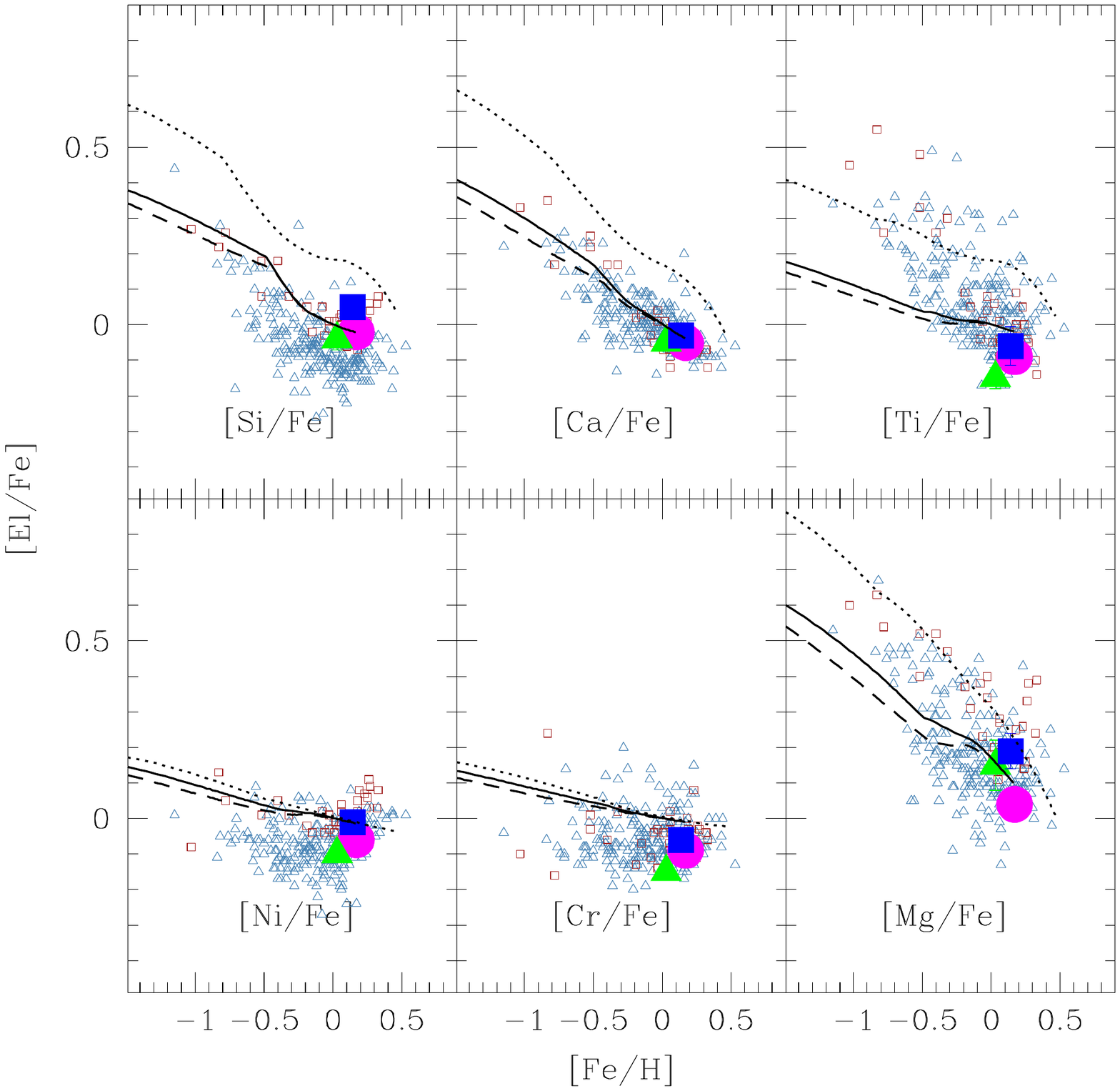}} 
   \hspace{5mm}
\caption{Abundance ratios in field and open cluster stars: the abundance ratios [El/Fe] versus [Fe/H] of solar neighbourhood dwarf stars (cyan empty triangles), of inner disk/bulge giant stars (red empty squares), of stars in Trumpler~20 (magenta filled  circle), 
in NGC~4815 (green filled triangle), and NGC~6705 (blue filled square). In panel a) the curves are the model of R10 at three R$_{\rm GC}$: 4~kpc (dotted line), 6~kpc (continuous  line), and 8~kpc (dashed line).  The curves shown in panel b) are obtained  with the model of M09. 
The theoretical ratios are normalised to the solar abundances predicted by each model, with the exception of [Mg/Fe] in M09 model, which is displaced by other $+$0.15 dex to match the data. }
\label{Fig_model1}
\end{figure*}

In Fig.\ref{Fig_model1} we show the results of the two models: R10 in panel a) and M09 in panel b). The curves refer to different Galactic radii, 4~kpc, 6~kpc and 8~kpc. Both models are those originally published by R10 and M09, and are not modified to reproduce the Gaia-ESO Survey data. 
Thus the agreement with the data, and the similarity of  their predictions for the Solar radius are encouraging, even if there are some discrepant elements. 
The only exception is Ni for which different yields are adopted by R10 and M09.  
We note also that the stellar yields adopted in both models  are not able to reproduce the trends of [Ti/Fe] vs [Fe/H], as already 
noticed in the original papers. 

Notwithstanding the agreement at the Solar radius, the two models show a different behaviour in the inner disk, where different assumptions of the infall and star formation rates are made.  
This might be ascribed to the paucity of observational constraints in the inner disk when the two models were designed. 
We note indeed that in panel a),  the model curves reach higher metallicities (by $\sim$0.2 dex) at the innermost radius because of the more intense SFR, but the predicted [El/Fe] ratios are pretty much the same. 
In panel b),  the evolution of the $\alpha$- and iron-peak elements behaves differently with respect to the Galactocentric radius: while the iron-peak elements are quite insensitive to the choice of the Galactocentric radius, the $\alpha$-elements are more enhanced in stars born at R$\sim$4~kpc than at the Solar radius. 
This can be easily explained with the short time scales for the gas consumption in the inner part of the disk 
due to the high star formation rate, which typically produce high  $\alpha$ over iron abundances. On the other hand, the iron-peak elements 
are insensitive to the  radius since they behave, to first order, as iron. 

For some $\alpha$-elements, Si and  Mg, and marginally also Ca and Ti, 
both the inner-disk/bulge stars and NGC~6705 are in 
better agreement with the curves corresponding to the innermost radii of M09, i.e. in between 4 and 6~kpc from the Galactic Center. 
On the other hand, the $\alpha$ abundance ratios of  Trumpler~20 and of NGC~4815 agree better with the curve corresponding 
to their  present radius in M09 and  with the curves of R10.
For the iron-peak elements the behaviour of the three clusters is very similar and in agreement with the model curves, of both M09 and R10, 
which do not predict strong variation of these ratios with the Galactocentric radii.

The comparison of the abundance ratios with the model curves together with the similarity observed between the cumulative distributions of abundance ratios and 
of the solar neighbourhood and inner-disk/bulge stars, gives us  a "chemical" indication of the birthplace of the 
three inner-disk clusters under analysis: 
the good agreement of the  [$\alpha$/Fe] abundance ratio of Trumpler~20 and NGC~4815 with the curves at $\sim$6-8~kpc of both models, and at the same time 
with the abundance ratios of solar neighbourhood stars,  
is a reasonable  indication that they were born not very far from their present location. 
On the other hand, the good agreement within the error of some abundance ratios of  NGC~6705 with M09 curves for radii $\sim$4-6~kpc, and with the observations of  inner-disk/bulge stars (see 
Fig.~\ref{fig_ks_6705}), might indicate that it has moved towards its present position from an inner birthplace. 
This is in agreement with the orbit determination done for this cluster in \citet{magrini10} with a perigalacticon of $\sim$5~kpc and apogalacticon of $\sim$~9~kpc. 
Under this scenario, a revision  of the basic assumptions of the model of 
R10 will be required for the model to be able to predict higher SFRs/steeper abundance gradients in the inner disk and to face the Gaia-ESO Survey abundance data for those clusters. For instance, the inclusion of a bar in the model could lead to enhanced star formation in the inner Galactic region \citep[see, e.g.,][] {wang12}.
In addition, the effect of local inhomogeneities, radial flows, stellar migration, and outflow   
should not be neglected in next generation of chemical evolution models.

\section{Summary}

In this  paper we present the analysis of abundance ratios in open clusters and field stars 
obtained in the first six months of the Gaia-ESO Survey. 
We studied three old/intermediate-age open clusters: NGC~6705, NGC~4815, and Trumpler~20. 
For the three clusters we find that: {\em i)} the clusters are internally homogeneous in the considered elements (four $\alpha$-elements, 
Si, Ca, Mg, Ti and three iron-peak elements, Fe, Ni, Cr); {\em ii)} the three clusters have similar [El/Fe] abundance patterns, but different global metallicity and, consequently, different 
[El/H] patterns; 
{\em iii)} a comparison of the cumulative distributions of abundance ratios shows that the abundance ratios of NGC~6705 are very similar to those of inner-disk/bulge stars studied by the Gaia-ESO Survey, while the abundance patterns of NGC~4815 and Trumpler~20 do not match perfectly either with the solar neighbourhood stars or with the inner-disk/bulge stars.  
We finally compare the field and cluster abundance ratios with two chemical evolution models (M09 and R10), finding a general 
good agreement for the solar neighbourhood. The predictions of the models differ 
for the inner disk. The $\alpha$-enhancement of NGC~6705 places it in better agreement with the 
model curves of M09 for Galactocentric radii from 4 to 6~kpc, and, together with its better agreement with the distributions of abundances ratios in the inner-disk/bulge sample, 
supports an inner birthplace for it. 
Concluding, the first results from the Gaia-ESO Survey show its huge potential to  give new constraints to our view of the Galactic chemical evolution, 
to explore areas of our Galaxy so far little studied, and moreover to put, for the first time, many stellar populations on exactly the same scale.

\begin{acknowledgements}
We acknowledge the support from INAF and Ministero dell'Istruzione, dell'Universit\'a e della Ricerca (MIUR) in the form of the grant "Premiale VLT 2012".
The results presented here benefited from discussions in three Gaia-ESO workshops supported by the ESF (European Science Foundation) through the GREAT (Gaia Research for European Astronomy Training) Research Network Program (Science meetings 3855, 4127 and 4415)
This work was partially supported by the Gaia Research for European Astronomy Training (GREAT-ITN) Marie Curie network, funded through the European Union Seventh Framework Programme [FP7/2007-2013] under grant agreement n. 264895. 
T.B. was funded by grant No. 621-2009-3911 from The Swedish Research Council.
This work was partly supported by the European Union FP7 programme through ERC grant number 320360. 
This work was partly supported by the Leverhulme Trust through grant RPG-2012-541.
I.S.R. gratefully acknowledges the support provided by the Gemini-CONICYT project 32110029.
This research has made use of the SIMBAD database, operated at CDS, Strasbourg, France.  
\end{acknowledgements}

\end{document}